\documentclass[prd,
 reprint,
 amsmath,amssymb,
 aps,
]{revtex4-1}

\usepackage{graphicx}
\usepackage{dcolumn}
\usepackage{bm}

\newcommand*{\bfrac}[2]{\genfrac{}{}{0pt}{}{#1}{#2}}

\begin{document}

\title{Quasinormal modes of Dirac spinors in the background of rotating black holes in four and five dimensions}

\author{Jose Luis Bl\'azquez-Salcedo}
\email[]{jose.blazquez.salcedo@uni-oldenburg.de}
\author{Christian Knoll}
\email[]{christian.knoll@uni-oldenburg.de}
\affiliation{Institut f\"ur Physik, Universit\"at Oldenburg, D-26111 Oldenburg, Germany}

\date{\today}

\begin{abstract}
We study the quasinormal modes of massive Dirac spinors in
the background of rotating black holes. 
In particular, we consider
the Kerr geometry 
as well as the five dimensional Myers-Perry spacetime 
with equal angular momenta. 
We decouple the equations using the standard methods from the literature. 
In the five dimensional Myers-Perry black hole the angular equation is solved analytically.
Using the continued fraction method, we calculate the spectrum of quasinormal modes 
for the ground modes and first excited modes.
We analyze, in a systematic way, its dependence on the different parameters of the black hole and fermionic field.
We compare our values with previous results available in the literature for Kerr and for the static limit.
The numerical results show several differences between the four and five dimensional cases. For instance, in five dimensions the symmetry between the positive and negative (real) frequency of the modes breaks down, which results in a richer spectrum.
\end{abstract}

\maketitle

\section{Introduction}
%
%
The study of black holes in higher dimensions \cite{Horowitz:2012nnc, Emparan:2008eg} has received a lot of attention in the last decades, due to several theories involving the existence of extra spatial dimensions, such as String theory, the Brane world theory and the AdS/CFT correspondence. 
Indeed, black holes can possess very different properties in more than four dimensions. For instance, when the dimension is higher than four they can have more than one angular momentum, and the topology of stationary event horizons can be different from spherical. Test fields in the black hole background can have very different properties too.

In this context it is of interest to study the properties of several fundamental fields as test fields in the black hole backgrounds, such as scalars, vectors or spinor fields, which present different features depending on the dimension.
 
Of special importance is the study of the behaviour of test fields in these geometries, in particular, of their quasinormal modes \cite{Konoplya:2011qq}. 
The quasinormal mode spectrum can be used to reveal several properties, such as instabilities, zero modes, bound states, or simply the resonances for which perturbations of these field tend to be radiated away and decay. The spectrum can help recognize possible backreacting configurations with non-trivial hair \cite{Blazquez-Salcedo:2018jnn}, and even if they are formally unstable, the order of magnitude of the instability can be such that the configuration could still have some physical relevance \cite{Degollado:2018ypf}.

In this paper we will study the quasinormal modes of the massive Dirac field in the five dimensional Myers-Perry black hole spacetime and compare to the behaviour of the quasinormal modes in the Kerr spacetime. This work will extend the previous results in the Tangherlini/Schwarzschild spacetimes, considered in four to nine dimensions \cite{Blazquez-Salcedo:2017bld}, and the results for the four dimensional Kerr-Newman background \cite{Konoplya:2017tvu}. We will use equal angular momenta for the Myers-Perry black hole due to the stronger resemblance to spherical symmetry.

The study of test fields and particles in the background of a Myers-Perry black hole is simplified by the separability of the relevant equations. For example the geodesic equations were studied in the five dimensional Myers-Perry black hole by a separation of the Hamilton-Jacobi equations in \cite{Frolov:2003en}. The massive scalar field equations in the same background were studied in \cite{Frolov:2002xf}. 
The Dirac equation can similarly be separated in these geometries, reducing the problem to ordinary differential equations.
The Ansatz by Chandrasekhar in the Kerr geometry \cite{Chandrasekhar:1984siy} was extended to the five dimensional Myers-Perry spacetime \cite{Wu:2008df}, and for the massive Dirac equation in the Kerr-NUT-AdS spacetimes in all dimensions \cite{Oota:2007vx, Cariglia:2011qb}. These results were also extended to the Proca field \cite{Lunin:2017drx}. The separation is based on the existence of a Killing-Yano tensor, which was constructed for the most general charged rotating geometries in \cite{Chervonyi:2015ima}. For more details we refer the reader to \cite{Frolov:2017kze}.

%
%

The spectrum of quasinormal modes of rotating black holes has been very well studied. 
The quasinormal modes of vector and tensor metric perturbations in the background of the Kerr black hole were calculated in \cite{Onozawa:1996ux} using the continued fraction method. The results where later improved in \cite{Cook:2014cta}. The spectrum is also known for the asymptotically AdS Kerr black hole \cite{Wang:2015fgp}. For scalar field perturbations in the Kerr background, the quasinormal modes were calculated in \cite{Glampedakis:2003dn}. Interestingly, both cases reveal a looping behaviour of the quasinormal modes in the complex plane for certain values of the parameters. 

Concerning higher dimensions, the quasinormal modes of a massless scalar field in the ultra-spinning limit of the six dimensional Myers-Perry black hole were studied in \cite{Morisawa:2004fs} using the continued fraction method too. The quasinormal modes of tensor metric perturbations were studied for odd dimensional Myers-Perry-AdS with equal angular momenta \cite{Kunduri:2006qa}, for Myers-Perry black holes with only one non vanishing angular momentum and dimension greater or equal to seven \cite{Kodama:2009bf} and for Myers-Perry-AdS with only one non-vanishing angular momentum \cite{Kodama:2009rq}. For more general Myers-Perry black holes, with two non-vanishing and independent angular momenta, the scalar field quasinormal modes were studied in \cite{Cho:2011yp}. A discussion of the massless scalar, vector and tensor quasinormal modes for different cases of the Myers-Perry black holes can be found in \cite{Dias:2014eua}.

The quasinormal modes of a massive vector field in the Myers-Perry-NUT-AdS background were calculated in \cite{Frolov:2018ezx}. It was shown that for certain mass values the spectrum develops quasiresonances (modes with very small damping time) and again a looping behaviour for the modes in the complex plane for certain values of the parameters.
%

%
%

Regarding Dirac test fields, the quasinormal modes for a massive Dirac field in the background of the Schwarzschild black hole were studied in \cite{Cho:2003qe} using the WKB method. Although only small mass values were considered, they were able to determine that an increase in mass leads generally to a decrease in the damping time and an increase in the real part of the frequency. The massless Dirac quasinormal modes of a Schwarzschild black hole were studied in \cite{Jing:2005dt} using the continued fraction and Hill-determinant methods. The quasinormal modes of a massless Dirac field in the spacetime of a spherically symmetric charged black hole in higher dimensions were studied in \cite{Chakrabarti:2008xz} using the WKB method. Similarly, in \cite{Saleh:2016pke} the quasinormal modes of the massless Dirac field and gravity perturbations of a quantum corrected Schwarzschild black hole were discussed. In  \cite{Fernando:2015hma} the quasinormal modes of the massless Dirac field in massive gravity were investigated, using in addition the P\"oschl-Teller approximation. 

As already mentioned, the quasinormal modes of the massive Dirac field in the background of the Kerr-Newman black hole were recently investigated in \cite{Konoplya:2017tvu}. They found quasiresonances (long-lived modes) and a fine structure depending on the angular quantum numbers. Similar modes were recently studied for the static Tangherlini space-time in \cite{Blazquez-Salcedo:2017bld}, finding similar quasiresonances. In addition, in \cite{Chowdhury:2018izv} the quasinormal modes of a charged, massless Dirac field and a charged, massive scalar field in the background of a spherically symmetric black hole with scalar hair in four dimensions have been studied using the continued fraction method. The massive Dirac field in the near-horizon region of the five dimensional Myers-Perry black hole with equal angular momenta was studied in \cite{Blazquez-Salcedo:2018rec}.

%
%

The structure of the paper is the following. In section \ref{sec:Kerr} we will decouple the massive Dirac equation in the Kerr spacetime. For this we will use the method from the literature which is deeply linked to the existence of a Killing-Yano tensor in the Kerr spacetime. The separation will allow us to derive the previously known recurrence relation obeyed by the angular quantum numbers, and to write down the coupled first order differential equations for the radial part of the spinor. In section \ref{sec:5d_MP} we will repeat the process for the five dimensional Myers-Perry black hole spacetime with equal angular momenta. The only difference being that we can solve the angular equation analytically and thus have an explicit formula of the angular eigenvalue. In section \ref{sec:method} we will discuss the numerical method used to calculate the quasinormal modes. After this we will show and discuss the obtained numerical results in section \ref{sec:results}. Lastly in section \ref{sec:conclusions} we wrap the paper up with a discussion of the results and possible future directions.

\section{Dirac spinors in the Kerr black hole background}\label{sec:Kerr}

The Kerr metric \cite{PhysRevLett.11.237} in Boyer-Lindquist coordinates \cite{doi:10.1063/1.1705193} is given by 
\begin{eqnarray}
\mathrm d s^2 &=& \left( \frac{\Delta}{\rho} [\mathrm d t + a \sin^2 \mathrm d \phi] \right)^2 - \left( \frac{\rho}{\Delta} \mathrm d r \right)^2 - \left( \rho \mathrm d \theta \right)^2 \nonumber \\
&&- \left( \frac{\sin \theta}{\rho} [R^2 \mathrm d \phi + a \mathrm d t] \right)^2 \, , 
\end{eqnarray}
with
\begin{eqnarray}
\rho^2 &=& r^2 + a^2 \cos^2 \theta \, , \nonumber \\
R^2 &=& r^2 + a^2 \, , \nonumber \\
\Delta^2 &=& r^2 - \mu r + a^2 \, . 
\end{eqnarray}
The parameter $\mu$ is related to the mass of the black hole, and $a$ is related to the angular momentum. We use the vielbein
\begin{eqnarray}\label{eqn:Kerr_basis}
\boldsymbol{\omega}^0 &=& \frac{\Delta}{\rho} (\mathbf d t + a \sin^2 \theta \, \mathbf d \phi) \, , \nonumber \\
\boldsymbol{\omega}^1 &=& \frac{\rho}{\Delta} \mathbf d r \, , \nonumber \\
\boldsymbol{\omega}^2 &=& \rho \mathbf d \theta \, , \nonumber \\
\boldsymbol{\omega}^3 &=& \frac{\sin \theta}{\rho} ( R^2 \mathbf d \phi + a \mathbf d t) \, .
\end{eqnarray}

This gives the Dirac operator
\begin{eqnarray}
\mathcal D &=& \frac{\mathrm i}{\Delta \rho} \gamma^0 (R^2 \partial_t - a \partial_\phi) + \frac{\mathrm i \Delta}{\rho} \gamma^1 \Bigg\{ \partial_r + \partial_r \ln \sqrt{\Delta \rho^2} \nonumber \\
&&- \frac{1}{2} \partial_r \arctan \left( \frac{a \cos \theta}{r} \right) \, \gamma^0 \gamma^1 \gamma^2 \gamma^3 \Bigg\} \nonumber \\
&&+ \frac{\mathrm i}{\rho} \gamma^2 \Bigg\{ \partial_\theta + \partial_\theta \ln \sqrt{\rho^2 \sin \theta} \nonumber \\
&&- \frac{1}{2} \partial_\theta \arctan \left( \frac{a \cos \theta}{r} \right) \gamma^0 \gamma^1 \gamma^2 \gamma^3 \Bigg\} \nonumber \\
&&+ \frac{\mathrm i}{\rho \sin \theta} \gamma^3 (\partial_\phi - a \sin^2 \theta \, \partial_t) \, .
\end{eqnarray}
To decouple this operator into radial and angular parts, we use the transformation \cite{Chandrasekhar:579245}
\begin{eqnarray}
\frac{1}{\zeta} \mathrm e^{\xi \, \gamma^0 \gamma^1 \gamma^2 \gamma^3} \left[ \mathcal D - m \right] \zeta \mathrm e^{\xi \, \gamma^0 \gamma^1 \gamma^2 \gamma^3} = \frac{1}{\rho} \mathcal D_\ast \, ,
\end{eqnarray}
with
\begin{eqnarray}
\zeta &=& \frac{1}{\sqrt{\Delta \rho^2 \sin \theta}} \, , \nonumber \\
\xi &=& \frac{1}{2} \arctan \left( \frac{a \cos \theta}{r} \right) \, .
\label{trans_kerr}
\end{eqnarray}

Before continuing, let us comment here that the principal conformal Killing-Yano tensor for the Kerr metric in the vielbein basis given by equation (\ref{eqn:Kerr_basis}) is \cite{doi:10.1111/j.1749-6632.1973.tb41447.x, Floyd}
\begin{eqnarray}
\mathbf h = r \boldsymbol{\omega}^1 \wedge \boldsymbol{\omega}^0 + a \cos \theta \, \boldsymbol{\omega}^2 \wedge \boldsymbol{\omega}^3 \, .
\label{KY_kerr}
\end{eqnarray}
Hence note that the argument of the $\arctan$-function in the transformation (\ref{trans_kerr}) can be expressed as the ratio of the two eigenvalues of the principal conformal Killing Yano tensor. 

The operator $\mathcal D_\ast$ can be written like
\begin{eqnarray}
\mathcal D_\ast = \frac{\mathrm i}{\Delta} \gamma^0 (R^2 \, \partial_t - a \, \partial_\phi) + \mathrm i \Delta \partial_r + \mathrm i \gamma^0 \gamma^1 \mathcal K_\ast - r m \, ,
\end{eqnarray}
with the angular operator
\begin{eqnarray}
\mathcal K_\ast &=& \hat \gamma^1 \partial_\theta + \frac{1}{\sin \theta} \hat \gamma^2 ( \partial_\phi - a \sin^2 \theta \, \partial_t ) \nonumber \\
&&+ \mathrm i m a \cos \theta \, \hat \gamma^1 \hat \gamma^2 \, ,
\end{eqnarray}
where we have defined $\hat \gamma^{j-1} = \gamma^0 \gamma^1 \gamma^j$ with $j \in \{2, 3\}$. 

The matrices $\gamma^0$ and $\gamma^1$ commute with the matrices $\hat \gamma^1$ and $\hat \gamma^2$ by definition and thus $[\mathcal D_\ast, \mathcal K_\ast] = 0$. With this we can split the spinor $\Psi$ in parts
\begin{eqnarray}
\frac{1}{\zeta} \mathrm e^{-\xi \gamma^0 \gamma^1 \gamma^2 \gamma^3} \Psi = \Psi_\ast = \phi(r) \otimes \Theta_\kappa (\theta) \, \mathrm e^{- \mathrm i \omega t + \mathrm i \lambda \phi} \, ,
\end{eqnarray}
with
\begin{eqnarray}
\mathcal K_\ast \Theta_\kappa (\theta) \, \mathrm e^{- \mathrm i \omega t + \mathrm i \lambda \phi} = \kappa \Theta_\kappa (\theta) \, \mathrm e^{- \mathrm i \omega t + \mathrm i \lambda \phi} \, .
\end{eqnarray}

It is convenient to introduce quantities scaled to the mass of the black hole:
\begin{eqnarray}
r = \mu x \, , 
a = \mu \alpha \, , 
\Omega = \mu \omega \, , 
\eta = \mu m \, .
\end{eqnarray}
Then the radial equation is given by 
\begin{eqnarray}
\left\{ (X^2 \Omega + \alpha \lambda) \gamma^0 + \mathrm i \Delta_x^2 \gamma^1 \frac{\mathrm d}{\mathrm d x} + \mathrm i \kappa \Delta_x \gamma^0 \gamma^1 - x \eta \Delta_x  \right\} \phi = 0 \nonumber \\
\end{eqnarray}
and the angular equation by
\begin{eqnarray}
\left\{ \hat \gamma^1 \frac{\mathrm d}{\mathrm d \theta} + \frac{\mathrm i (\lambda + \alpha \Omega \sin^2 \theta)}{\sin \theta} \hat \gamma^2 + \mathrm i \alpha \eta \cos \theta \hat \gamma^1 \hat \gamma^2 - \kappa \right\} \Theta_\kappa = 0 \, , \nonumber \\
\end{eqnarray}
with
\begin{eqnarray}
X^2 &=& x^2 + \alpha^2 \, , \nonumber \\
\Delta_x^2 &=& x^2 - x + \alpha^2 = (x-x_+)(x-x_-) \, , \nonumber \\
x_\pm &=& \frac{1 \pm \sqrt{1 - 4 \alpha^2}}{2} \, .
\end{eqnarray}

\subsection{The angular equation in the Kerr background}

Following \cite{doi:10.1063/1.529963, doi:10.1063/1.525820, Konoplya:2017tvu} we calculate the angular eigenvalue numerically with a continued fraction stemming from a recurrence relation coming from an expansion of the angular function in hypergeometric functions. For this we change the variables to $y = (1 + \mathrm{sgn}(\lambda) \cos \theta) / 2 \in [0, 1]$, choose the $\gamma$ matrices representation to be
\begin{eqnarray}
\hat \gamma^1 = \left[ \begin{array}{cc} 0 & 1 \\ -1 & 0 \end{array} \right] \; , \; \hat \gamma^2 = \left[ \begin{array}{cc} 0 & \mathrm i \\ \mathrm i & 0 \end{array} \right]  \; , \; \Theta_\kappa = \left[ \begin{array}{c} \Theta_1 \\ \Theta_2 \end{array} \right]  \, ,
\end{eqnarray}
and factorize the $\Theta$ functions like
\begin{eqnarray}
\Theta_1 =& y^{|\lambda|/2} (1-y)^{(|\lambda|+1)/2} h_1 \, , \nonumber \\
\Theta_2 =& y^{(|\lambda| + 1)/2} (1-y)^{|\lambda|/2} h_2 \, . 
\end{eqnarray}

The equations fulfilled by the functions $h_1$ and $h_2$ then read
\begin{eqnarray}
\left[ p - \Sigma (1-y) - (1-y) \frac{\mathrm d}{\mathrm d y} \right] h_1 &=& - [\sigma (2 y - 1 ) + \tilde \kappa] h_2 \, , \nonumber \\
\left[ p - \Sigma y + y \frac{\mathrm d}{\mathrm d y} \right] h_2 &=& + [\sigma (2 y - 1) - \tilde \kappa] h_1 \, , \nonumber \\
\end{eqnarray}
with
\begin{eqnarray}
p &:=& |\lambda| + \frac{1}{2} \, , \nonumber \\
\Sigma &:=& -\mathrm{sgn}(\lambda) 2 \alpha \Omega \, , \nonumber \\
\tilde \kappa &:=& \mathrm{sgn}(\lambda) \kappa \, , \nonumber \\
\sigma &:=& -\alpha \eta \, .
\end{eqnarray}
We now use a series expansion
\begin{eqnarray}
h_1 &=& \sum\limits_{n=0}^\infty c_n \mathcal F^-_n \, , \nonumber \\
h_2 &=& - \mathrm{sgn}(\lambda) \epsilon_\kappa \sum\limits_{n=0}^\infty (-1)^n \frac{n + p}{p} c_n \mathcal F^+_n \, ,
\label{ser_ex}
\end{eqnarray}
with $\epsilon_\kappa = \pm 1$ being a sign choice and
\begin{eqnarray}
\mathcal F^\pm_n := F \left( \bfrac{-n, n + 2 p}{p + \frac{1 \pm 1}{2}} ; y \right) 
\end{eqnarray}
being hypergeometric functions. Making use of the special relations between contiguous hypergeometric functions \cite[section 15.5(ii)]{NIST:DLMF}, 
it is possible to rewrite the differential equations for $h_1$ and $h_2$ into a
recurrence relation for the coefficients of the series expansion (\ref{ser_ex}):
\begin{eqnarray}\label{eqn:angular_recc}
&&\frac{n+2p-1}{2(n+p)-1} \left[ \sigma_n - \frac{\Sigma}{2} \right] c_{n-1} \nonumber \\
&&+ \left[ n + p + \kappa_n + \frac{(2p -1) (\sigma_n - \Sigma (n+p) )}{(2[n+p]-1)(2[n+p]+1)}  \right] c_n \nonumber \\
&&+ \frac{n+1}{2(n+p)+1} \left[\sigma_n + \frac{\Sigma}{2} \right] c_{n+1} = 0 \, ,
\end{eqnarray}
with $\sigma_n = - \epsilon_\kappa \mathrm{sgn}(\lambda) (-1)^n$ and $\kappa_n = - \epsilon_\kappa \mathrm{sgn}(\lambda) (-1)^n \kappa$. Defining $\Delta_n = c_{n} / c_{n+1}$ this can be written as a continued fraction \cite{Leaver:1985ax}. Given $\Omega$, $\epsilon_\kappa$ and $\lambda$, we can search in the complex plane for the corresponding value of $\kappa$ by evaluating this continued fraction $\Delta_n$ to a finite depth requiring that it converges to the value given by the original recurrence relation evaluated at $n=0$, using $c_{-1} = 0$ \cite{Konoplya:2017tvu} and
\begin{eqnarray}\label{eqn:Kerr_angular_cfe}
\left[  p + \kappa_0 + \frac{\sigma_0 - \Sigma p}{2 p + 1}  \right] \Delta_1 + \frac{1}{2 p +1} \left[\sigma_0 + \frac{\Sigma}{2} \right] = 0 \, .
\end{eqnarray}

\subsection{The radial equation in the Kerr background}

We choose the following representation for the matrices $\gamma^0, \gamma^1$
\begin{eqnarray}
\gamma^0 = \left[ \begin{array}{cc} 0 & 1 \\ 1 & 0 \end{array} \right] \; , \; \gamma^1 = \left[ \begin{array}{cc} 0 & 1 \\ -1 & 0 \end{array} \right] \; , \; \phi = \left[ \begin{array}{c} \phi_{1} \\ \phi_{2} \end{array} \right] \, .
\label{kerr_rad_rep}
\end{eqnarray}

Since we are interested in perturbations whose flux falls into the black hole horizon, the boundary condition $\overline{\Psi} \gamma^1 \Psi \le 0$ for $r \rightarrow r_+$ gives the asymptotic behaviour
\begin{eqnarray}
\left[ \begin{array}{c} \phi_1 \\ \phi_2 \end{array} \right] \sim \left( \frac{x - x_+}{x - x_-} \right)^\beta \left[ \begin{array}{c} \phi_{1, r_+} \\ \sqrt{ \frac{x - x_+}{x - x_-} } \, \phi_{2, r_+} \end{array} \right] \, , 
\end{eqnarray}
with
\begin{eqnarray}
\beta &=& - \frac{\mathrm i (\alpha \lambda + [x_+^2 + \alpha^2] \Omega)}{x_+ - x_-} \nonumber \\
&=& - \mathrm i \left. \frac{X^2 \Omega + \alpha \lambda}{\partial_x (\Delta_x^2)} \right|_{x=x_+} \, , \nonumber \\
\phi_{1, r_+} &=& \mathrm i (x_+ - x_-) \left( 2 \beta + \frac{1}{2} \right) \, , \nonumber \\
\phi_{2, r_+} &=& \sqrt{x_+ - x_-} [\eta x_+ + \mathrm i \kappa]
\end{eqnarray}

At infinity, we are interested in perturbations whose flux is radiated away from the black hole. Hence we have $\overline{\Psi} \gamma^1 \Psi \ge 0$ for $r \rightarrow \infty$, which gives the asymptotic behaviour
\begin{eqnarray}
\left[ \begin{array}{c} \phi_1 \\ \phi_2 \end{array} \right] \sim (x-x_-)^{\tilde \beta} \mathrm e^{\mathrm i \chi x} \left[ \begin{array}{c} \Omega - \chi \\ \eta \end{array} \right] \, ,
\end{eqnarray}
with $\chi = \sqrt{\Omega^2 - \eta^2}$, $\mathrm{sgn} (\Omega) = \mathrm{sgn} (\chi)$ and
\begin{eqnarray}
\tilde \beta = \frac{\mathrm i}{2} \frac{2 \Omega^2 - \eta^2}{\chi} \, .
\end{eqnarray}

We now factorize both of these asymptotic behaviours from the spinor
\begin{eqnarray}
\left[ \begin{array}{c} \phi_1 \\ \phi_2 \end{array} \right] = \left( \frac{x - x_+}{x - x_-} \right)^\beta (x-x_-)^{\tilde \beta} \mathrm e^{\mathrm i \chi x} \left[ \begin{array}{c} \psi_1 \\ \sqrt{\frac{x-x_+}{x-x_-}} \psi_2 \end{array} \right] \, .
\end{eqnarray}
We also change variables to
\begin{eqnarray}
y = \frac{x - x_+}{x - x_-} \, .
\label{def_y}
\end{eqnarray}
This gives the following system of first order differential equations for $\psi_1$ and $\psi_2$
\begin{eqnarray}\label{eqn:radial_system_d=4}
\left( S_1 - \mathrm i (1-y)^2 \frac{\mathrm d}{\mathrm d y} \right) \psi_1 &=& P_- \psi_2 \, , \nonumber \\
\left( S_2 + \mathrm i y (1-y)^2 \frac{\mathrm d}{\mathrm d y} \right) \psi_2 &=& P_+ \psi_1 \, ,
\end{eqnarray}
with
\begin{eqnarray}
P_\pm &=& \eta x_+ \pm \mathrm i \kappa - [\eta x_- \pm \mathrm i \kappa] y \, , \nonumber \\
S_1 &=& \Omega (2 x_+ - y ) - i \tilde \beta (1-y) + \chi (x_+ -  x_-)  \, , \nonumber \\
 S_2 &=&  \frac{\Omega Y^2 + Q (1-y)^2}{x_+ - x_-} - y \left[ - \mathrm i \tilde \beta (1-y) +\chi (x_+ -  x_-) \right] \, , \nonumber \\
Q &=& \alpha \lambda + \mathrm i \left[ \beta + \frac{1}{2} \right] (x_+ - x_-) \, .
\end{eqnarray}

\section{Dirac Spinors in the Five dimensional Myers Perry black hole background}\label{sec:5d_MP}

Now that we have the angular and radial equations for the Dirac spinors in Kerr, let us move to the five dimensional case. We will repeat the previous steps for the five dimensional Myers-Perry black hole with equal angular momenta \cite{Myers:1986un}. The metric can be written like
\begin{eqnarray}
\mathrm d s^2 = \mathrm d t^2 - \frac{\mu^2}{R^2} \left( \mathrm d t + a \sin^2 \theta \, \mathrm d \phi_1 + a \cos^2 \theta \, \mathrm d \phi_2 \right)^2 \nonumber \\ 
 - \frac{r^2 R^2}{\Delta^2} \mathrm d r^2 - R^2 \left( \mathrm d \theta^2 + \sin^2 \theta \, \mathrm d \phi_1^2 + \cos^2 \theta \, \mathrm d \phi_2^2 \right) \, ,
\end{eqnarray}
where again $\mu$ is related to the mass of the black hole and $a$ is related to the angular momentum. We have defined
\begin{eqnarray}
R^2 &=& r^2 + a^2 \, , \nonumber \\
\Delta^2 &=& R^4 - \mu^2 r^2 = (r^2-r^2_+)(r^2-r^2_-) \, , \nonumber \\
r_\pm &=& \frac{\mu}{2} \pm \frac{\sqrt{\mu^2 - 4 a^2}}{2} \, .
\end{eqnarray}
We use the vielbein
\begin{eqnarray}\label{eqn:5dMP_basis}
\boldsymbol{\omega}^0 &=& \frac{\Delta}{r R} \left( \mathbf d t + a \sin^2 \theta \, \mathbf d \phi_1 + a \cos^2 \theta \, \mathbf d \phi_2 \right) \, , \nonumber \\
\boldsymbol{\omega}^1 &=& \frac{r R}{\Delta} \, \mathbf d r \, , \nonumber \\
\boldsymbol{\omega}^2 &=& R \, \mathbf d \theta \, , \nonumber \\
\boldsymbol{\omega}^3 &=& R \cos \theta \, \sin \theta \, \left( \mathbf d \phi_1 - \mathbf d \phi_2 \right) \, ,\nonumber \\
\boldsymbol{\omega}^4 &=& \frac{1}{r} \left(a \mathbf d t + R^2 \sin^2 \theta \, \mathbf d \phi_1 + R^2 \cos^2 \theta \, \mathbf d \phi_2 \right) \, .
\end{eqnarray}
This gives the Dirac operator
\begin{eqnarray}
\mathcal D  &=& \frac{\mathrm i R}{\Delta r} \gamma^0 \left( R^2 \partial_t - a \left[ \partial_{\phi_1} + \partial_{\phi_2} \right] \right) \nonumber \\
&&+ \frac{\mathrm i \Delta}{r R} \gamma^1 \left( \partial_r + \partial_r \ln \sqrt{\Delta R} - \frac{1}{2} \partial_r \arctan \left[ \frac{a}{r} \right] \gamma^0 \gamma^1 \gamma^2 \gamma^3 \right) \nonumber \\
&&+ \frac{\mathrm i}{R} \gamma^2 \left( \partial_\theta + \partial_\theta \ln\sqrt{\cos \theta \, \sin \theta} \right) \nonumber \\
&&+ \frac{\mathrm i}{R} \gamma^3 \left( \cot \theta \, \partial_{\phi_1} - \tan \theta \, \partial_{\phi_2} \right) \nonumber \\
&&+ \frac{\mathrm i}{r} \gamma^4 \left( \partial_{\phi_1} + \partial_{\phi_2} - a \partial_t \right) \nonumber \\
&&- \frac{\mathrm i a}{2 r^2} \gamma^0 \gamma^1 \gamma^4 - \frac{\mathrm i}{2 r} \gamma^2 \gamma^3 \gamma^4  \, . 
\end{eqnarray}
To decouple the angular and the radial parts, we use the same method as in four dimensions \cite{Oota:2007vx, Wu:2008df, Cariglia:2011qb}, making use of the transformation
\begin{eqnarray}
\frac{1}{\zeta} \mathrm e^{\xi \gamma^0 \gamma^1 \gamma^2 \gamma^3} \left( \mathcal D - m  \right) \zeta \mathrm e^{\xi \gamma^0 \gamma^1 \gamma^2 \gamma^3} =:  \mathcal D_\ast \, ,
\end{eqnarray}
with
\begin{eqnarray}
\zeta &=& \frac{1}{\sqrt{\Delta R \sin \theta \, \cos \theta}} \, , \nonumber \\
\xi &=& \frac{1}{2} \arctan \left( \frac{a}{r} \right) \, .
\end{eqnarray}

Let us compare these expressions with the four dimensional case. The principal conformal Killing-Yano tensor for the five dimensional Myers-Perry metric with equal angular momenta in the vielbein given by \ref{eqn:5dMP_basis} reads \cite{Frolov:2007nt}
\begin{eqnarray}
\mathbf h = r \boldsymbol{\omega}^1 \wedge \boldsymbol{\omega}^0 + a \boldsymbol{\omega}^2 \wedge \boldsymbol{\omega}^3 \, .
\end{eqnarray}
Note that, similiar to the Kerr case (\ref{KY_kerr}), the argument of the arctan-function in the above transformation can be expressed as the ratio of the two eigenvalues of the principal conformal Killing Yano tensor. 

The operator $\mathcal D_\ast$ can then be written as
\begin{eqnarray}
\mathcal D_\ast &=& \frac{\mathrm i R}{\Delta r} \gamma^0 \left( R^2 \partial_t - a \left[ \partial_{\phi_1} + \partial_{\phi_2} \right] \right) + \frac{\mathrm i \Delta}{r R} \gamma^1 \partial_r \nonumber \\
&&+ \frac{\mathrm i a}{r R} \left( \partial_{\phi_1} + \partial_{\phi_2} - a \partial_t \right) - \frac{\mathrm i a^2}{2 r^2 R} \gamma^0 \gamma^1   \nonumber \\
&& + \frac{i}{R} \gamma^0 \gamma^1 \mathcal K_\ast - \frac{m r}{R} \, ,
\end{eqnarray}
with
\begin{eqnarray}
\mathcal K_\ast &=& \hat \gamma^2 \partial_\theta + \hat \gamma^3 \left( \cot \theta \, \partial_{\phi_1} - \tan \theta \, \partial_{\phi_2} \right) \nonumber \\
&&+ \hat \gamma^4 \left( \partial_{\phi_1} + \partial_{\phi_2} - a \left[ \mathrm i m + \partial_t \right] \right) - \frac{1}{2} 
\end{eqnarray}
and $\hat \gamma^j := \gamma^0 \gamma^1 \gamma^j$ with $j \in \{2, 3, 4\}$. Since the matrices $\gamma^0$ and $\gamma^1$ commute with the matrices $\hat \gamma^j$, we also have that $[\mathcal D_\ast, \mathcal K_\ast] = 0$. We can separate the angular and radial equations with the following Ansatz for the spinor $\Psi$:
\begin{eqnarray}
&&\frac{1}{\zeta} \mathrm e^{-\xi \gamma^0 \gamma^1 \gamma^2 \gamma^3} \Psi = \Psi_\ast \nonumber \\
&&= \phi(r) \otimes \Theta_\kappa (\theta) \, \mathrm e^{- \mathrm i \omega t + \mathrm i m_1 \phi_1 + \mathrm i m_2 \phi_2} \, ,
\end{eqnarray}
with
\begin{eqnarray}
\mathcal K_\ast \Theta_\kappa (\theta) \, \mathrm e^{- \mathrm i \omega t + \mathrm i m_1 \phi_1 + \mathrm i m_2 \phi_2} \nonumber \\
= \kappa \Theta_\kappa (\theta) \, \mathrm e^{- \mathrm i \omega t + \mathrm i m_1 \phi_1 + \mathrm i m_2 \phi_2} \, .
\end{eqnarray}
Once more it is useful to work with quantities scaled to the mass of the black hole,
\begin{eqnarray}
r = \mu x \, , 
a = \mu \alpha \, , 
\Omega = \mu \omega \, , 
\eta = \mu m  \, .
\end{eqnarray}
The radial equation then reads
\begin{eqnarray}
&&\left\{ X^2 \Sigma \gamma^0 + \mathrm i \Delta^2 \gamma^1 \frac{\mathrm d}{\mathrm d x} \right. \nonumber \\
&&\left.+ \mathrm i \Delta \left( \kappa x - \frac{\alpha^2}{2 x} \right) \gamma^0 \gamma^1 - \Delta (\Sigma_0 + \eta x^2)  \right\} \phi = 0 \, ,
\end{eqnarray}
with
\begin{eqnarray}
X^2 &=& x^2 + \alpha^2 \, , \nonumber \\
\Sigma &=& X^2 \Omega + \alpha \lambda = x^2 \Omega + \Sigma_0 \, , \nonumber \\
\Delta &=& \sqrt{(x - x_+)(x-x_-)(x+x_+)(x+x_-)} \, , \nonumber \\
\Sigma_0 &=& \Sigma(x=0) \, , \nonumber \\
\lambda &=& m_1 + m_2 \, , \nonumber \\
x_\pm &=& \frac{1}{2} \pm \frac{\sqrt{1 - 4 \alpha^2}}{2}  \, .
\end{eqnarray}
The angular equation is
\begin{eqnarray}
&&\left\{ \hat \gamma^2 \frac{\mathrm d}{\mathrm d \theta} + \mathrm i \hat \gamma^3 (m_1 \cot \theta - m_2 \tan \theta) \right. \nonumber \\
&&\left. + \mathrm i \hat \gamma^4 (\lambda - \alpha \eta + \alpha \Omega) - \left( \frac{1}{2} + \kappa \right)  \right\} \Theta_\kappa = 0 \, .
\end{eqnarray}

Note that the magnetic quantum numbers $m_1$ and $m_2$ only appear explicitly in the radial equation in the combination $\alpha (m_1 + m_2) = \alpha \lambda$. 

Let us discuss the dependence of the radial equation on $\alpha$. Note that for $\lambda = 0$ the radial equation depends explicitly only on $\alpha^2$ and implicitly on $\kappa(\alpha)$.
Hence the modes will not depend on the 
sign of $\alpha$ 
 if $\kappa$ does not depend on the sign of $\alpha$ in the case of $\lambda = 0$.
We will show that this is the case in the next section.

\subsection{Solving the angular equation in the Myers-Perry background}

The angular equation in the five dimensional case turns out to be analytically solvable. Solutions have been already obtained in the near-horizon limit of the extremal case \cite{Blazquez-Salcedo:2018rec}. We use for the representation of the algebra fulfilled by the matrices $\hat \gamma^j$
\begin{eqnarray}
\hat \gamma^2 &=& \left[ \begin{array}{cc} 0 & 1 \\ -1 & 0 \end{array} \right] \; , \; \hat \gamma^3 = \left[ \begin{array}{cc} 0 & \mathrm i \\ \mathrm i & 0 \end{array} \right] \, , \nonumber \\
\hat \gamma^4 &=& - \hat \gamma^2 \hat \gamma^3 = \left[ \begin{array}{cc} - \mathrm i & 0 \\ 0 & \mathrm i \end{array} \right] \; , \; \Theta_\kappa = \left[ \begin{array}{c} \Theta_1 \\ \Theta_2 \end{array} \right] \, .
\end{eqnarray}
The system of differential equations is then
\begin{eqnarray}
\left( \frac{\mathrm d}{\mathrm d \theta} + m_1 \cot \theta - m_2 \tan \theta \right) \Theta_1 &=& - K_+ \Theta_2 \, , \nonumber \\
\left( \frac{\mathrm d}{\mathrm d \theta} - m_1 \cot \theta + m_2 \tan \theta \right) \Theta_2 &=& +K_- \Theta_1 \, ,
\end{eqnarray}
where we have defined
\begin{eqnarray}
K_\pm &:=& \frac{1}{2} + \kappa \pm [m_1 + m_2 + \alpha (\Omega - \eta)] \, .
\end{eqnarray}
Let us also define
\begin{eqnarray}
C &:=& \cos \theta \; , \; S := \sin \theta \, , \nonumber \\
p_1 &:=& \left| m_1 + \frac{1}{2} \right| \; , \; p_2 := \left| m_2 + \frac{1}{2} \right| \, , \nonumber \\
\mathcal F_j &:=& F \left( \bfrac{j + 1 - n_\kappa, j + n_\kappa + p_1 + p_2}{j + 1 + p_2} ; C^2 \right) \, , \nonumber \\
\mathcal R_j &:=& \frac{(j + 1 - n_\kappa) \left(j + n_\kappa + p_1 + p_2 \right)}{j + 1 + p_2} \, , 
\end{eqnarray}
where $F(a, b; c; z)$ is the hypergeometric function, $n_\kappa \ge 1$ is a natural number and $m_1$ and $m_2$ are half integer numbers. The solution in the case $K_+ \neq 0$ is (up to the normalization)
\begin{eqnarray}
\Theta_1 &=& C^{p_2 + 1/2} \, S^{p_1 + 1/2} \, \mathcal F_0 \, , \nonumber \\
\Theta_2 &=& \left\{ \frac{2 C S \, \mathcal R_0 \mathcal F_1}{\mathcal F_0} - \left( m_1 + \frac{1}{2} + p_1 \right) \frac{C}{S} \right. \, , \nonumber \\
&&\left. \;\; + \left( m_2 + \frac{1}{2} + p_2 \right) \frac{S}{C} \right\} \frac{\Theta_1}{K_+} 
\end{eqnarray}
and the angular eigenvalue in this case is
\begin{eqnarray}\label{eqn:MP_angular_EV_full}
\kappa &=& - \frac{1}{2} \pm_\kappa \sqrt{\Lambda + L } \, , \nonumber \\
\Lambda &=& \alpha (\Omega - \eta) \left[ \alpha (\Omega- \eta) + 2 \lambda \right]  \, , \nonumber \\
L &=& \left( 2 n_\kappa - 1 + p_1 + p_2 \right)^2 \, ,
\end{eqnarray}
where $\pm_\kappa$ is a sign choice. Note that in the case $\lambda = 0$ the above angular eigenvalue does not depend on the sign of $\alpha$.

In the case of $K_+ = 0$ the solution is (up to the normalization)
\begin{eqnarray}
\Theta_1 &=& 0 \, , \nonumber \\
\Theta_2 &=& C^{m_2} S^{m_1} \, ,
\end{eqnarray}
with $m_1, m_2 \ge 1/2$. The angular eigenvalue in this case is
\begin{eqnarray}\label{eqn:MP_angular_EV_special_1}
\kappa = -\frac{1}{2} - \lambda - \alpha (\Omega - \eta) \, .
\end{eqnarray}

Lastly in the case of $K_- = 0$ the solution is
\begin{eqnarray}
\Theta_1 &=& C^{-m_2} S^{-m_1} \, , \nonumber \\
\Theta_2 &=& 0 \, ,
\end{eqnarray}
with $m_1, m_2 \le - 1/2$. The angular eigenvalue in this case is
\begin{eqnarray}\label{eqn:MP_angular_EV_special_2}
\kappa = -\frac{1}{2} + \lambda + \alpha (\Omega - \eta) \, .
\end{eqnarray}

\subsection{The radial equation in the Myers-Perry black hole background}

For the radial part of the equation in Myers-Perry we choose the same representation for the matrices $\gamma^0, \gamma^1$ as in Kerr, given in equation (\ref{kerr_rad_rep}).
The same reasoning applies, concerning the boundary conditions. We want the flux to be infalling at the horizon, so $\overline{\Psi} \gamma^1 \Psi \le 0$ for $r \rightarrow r_+$ gives the asymptotic behaviour
\begin{eqnarray}
\left[ \begin{array}{c} \phi_1 \\ \phi_2 \end{array} \right] \sim \left( \frac{x - x_+}{x - x_-} \right)^\beta \left[ \begin{array}{c} \phi_{1, r_+} \\ \sqrt{ \frac{x - x_+}{x - x_-} } \, \phi_{2, r_+} \end{array} \right] \, , 
\end{eqnarray}
with
\begin{eqnarray}
\beta &=& - \frac{\mathrm i}{2} \frac{(x_+^2 + \alpha^2) \left[ (x_+^2 + \alpha^2) \Omega + \alpha \lambda \right]}{x_+ (x_+^2 - x_-^2)} \nonumber \\
&=& - \mathrm i \left. \frac{X^2 \Sigma}{\partial_x (\Delta^2)} \right|_{x=x_+} \, , \nonumber \\
\phi_{1, r_+} &=& \mathrm i x_+ (x_+^2 - x_-^2) \left( 3 \beta + \frac{1}{2} \right) \, , \nonumber \\
\phi_{2, r_+} &=& - \sqrt{2 x_+ (x_+^2 - x_-^2)} \nonumber \\
&&\left[ \Sigma_0 + \eta x_+^2 + \mathrm i \left( \kappa x_+ - \frac{\alpha^2}{2 x_+} \right) \right]
\end{eqnarray}
At infinity, we want an outgoing flux, so $\overline{\Psi} \gamma^1 \Psi \ge 0$ for $r \rightarrow \infty$ gives the asymptotic behaviour
\begin{eqnarray}
\left[ \begin{array}{c} \phi_1 \\ \phi_2 \end{array} \right] \sim \mathrm e^{\mathrm i \chi x} \left[ \begin{array}{c} \Omega - \chi \\ \eta \end{array} \right] \, ,
\end{eqnarray}
with $\chi = \sqrt{\Omega^2 - \eta^2}$ and $\mathrm{sgn} (\Omega) = \mathrm{sgn} (\chi)$.

We now factorize both asymptotic behaviours from the spinor
\begin{eqnarray}
\left[ \begin{array}{c} \phi_1 \\ \phi_2 \end{array} \right] = \left( \frac{x - x_+}{x - x_-} \right)^\beta \mathrm e^{\mathrm i \chi x} \left[ \begin{array}{c} \psi_1 \\ \sqrt{\frac{x-x_+}{x-x_-}} \psi_2 \end{array} \right] \, .
\end{eqnarray}
We also change to the $y$ variable defined in equation (\ref{def_y}) for Kerr.
This gives the following system of first order equations for $\psi_1$ and $\psi_2$
\begin{eqnarray}\label{eqn:radial_system_d=5}
\left( S_1 + \mathrm i \Delta_y \xi (1-y)^2 \frac{\mathrm d}{\mathrm d y} \right) \psi_1 &=& \sqrt{\Delta_y} P_- \psi_2 \, , \nonumber \\
\left( S_2 + \mathrm i \Delta_y \xi y (1-y)^2 \frac{\mathrm d}{\mathrm d y} \right) \psi_2 &=& \sqrt{\Delta_y} P_+ \psi_1 \, ,
\end{eqnarray}
with the definitions
\begin{eqnarray}
\xi &=& x_+ - x_- y \, , \nonumber \\
Y^2 &=& \xi^2 + \alpha^2 [1-y]^2 \, , \nonumber \\
\Sigma_y &=& \xi^2 \Omega + \Sigma_0 [1-y]^2 \, , \nonumber \\
\Delta_y &=& [2 x_+ - (x_+ + x_-) y][x_+ + x_- - 2 x_- y] \, , \nonumber \\
Q &=& (1-y)^2 \left[\beta + \frac{1}{2} \right] + \mathrm i \chi y (x_+ - x_-) \, , \nonumber \\
P_\pm &=& \mathrm i (1-y) ( \kappa \xi^2 - \alpha^2 (1-y)^2 / 2 ) \nonumber \\
&&\pm \xi (\Sigma_0 (1-y)^2 + \eta \xi^2 ) \, , \nonumber \\
R_1 &=& x_- (\xi + x_+) + 2 \alpha^2 - y \alpha^2 \, , \nonumber \\
R_2 &=& \Sigma_y + (x_+^2 + \alpha^2) \Omega \, , \nonumber \\
R_3 &=& x_+^2 - x_-^2 + 4 x_+ x_- - 2 x_- (x_+ - x_-) y \, , \nonumber \\
R_4 &=& \chi (x_+ - x_-) + \mathrm i \beta (2 - y) \, , \nonumber \\
S_1 &=& \xi \left[ \frac{R_1 R_2 + (2-y) \alpha \lambda}{x_+ - x_-} - \mathrm i \beta R_3 - \Delta_y R_4 \right] \, , \nonumber \\
S_2 &=& \xi \left[ \frac{Y^2 \Sigma_y}{x_+ - x_-} + \mathrm i \Delta_y Q \right]\, .
\end{eqnarray}

\section{Numerical method}\label{sec:method}

The first order systems (\ref{eqn:radial_system_d=4}) and (\ref{eqn:radial_system_d=5}) that describe the radial part of the Dirac spinor have the gestalt
\begin{eqnarray}
\left( B_1 + \Delta A_1 \frac{\mathrm d}{\mathrm d y} \right) \psi_1 &=& \sqrt{\Delta} C_1 \psi_2 \, \nonumber \, , \\
\left( B_2 + \Delta A_2 \frac{\mathrm d}{\mathrm d y} \right) \psi_2 &=& \sqrt{\Delta} C_2 \psi_1 \, ,
\end{eqnarray}
where $A_1, A_2, B_1, B_2, C_1, C_2$ and $\Delta$ are polynomials in $y$. We can rewrite this system as a second order equation for $\psi_1$
\begin{eqnarray}\label{eqn:second_order}
&& \Delta^2 A_1 A_2 C_1 \frac{\mathrm d^2 \psi_1}{\mathrm d y^2} 
+ \Delta \left\{ C_1 P_{1,2} - \Delta A_1 A_2 \frac{\mathrm d C_1}{\mathrm d y} \right\} \frac{\mathrm d \psi_1}{\mathrm d y}  \nonumber \\ 
&& + \left\{ C_1 \left[ B_1 B_2 - \Delta C_1 C_2 \right] + A_2 R_{1,2} \right\} \psi_1 = 0 \, ,
\end{eqnarray}
with
\begin{eqnarray}
P_{1,2} &=&  A_1 B_2 + A_2 B_1 + \frac{1}{2} A_1 A_2 \frac{\mathrm d \Delta}{\mathrm d y} + \Delta A_2 \frac{\mathrm d A_1}{\mathrm d y} \, , \nonumber \\
R_{1,2} &=& \Delta C_1 \frac{\mathrm d B_1}{\mathrm d y} - \frac{1}{2} B_1 C_1 \frac{\mathrm d \Delta}{\mathrm d y} - \Delta B_1 \frac{\mathrm d C_1}{\mathrm d y} \, .
\end{eqnarray}
Alternatively, we can write the second order differential equation for $\psi_2$, which is simply obtained by interchanging $1 \leftrightarrow 2$ in the above second order equation for $\psi_1$. 

Observe that all coefficients in the second order equation (\ref{eqn:second_order}) are polynomials in $y$. Thus the Frobenius method
\begin{eqnarray}\label{eqn:frobenius_ansatz}
\psi_{1, 2} = \sum\limits_{n=0}^\infty c^{(1, 2)}_n y^n
\end{eqnarray}
allows us to obtain a finite recurrence relation in the coefficients $c^{(1, 2)}_n$. This series converges on the unit circle and thus on the interval of interest as long as all the zeroes of the coefficient of the term $\psi_{1, 2}^{\prime \prime}$ are either at $0$ or have an absolute value $\ge 1$. These are the zeroes of $\Delta, A_1, A_2$ and $C_j$ for the second order equation of $\psi_j$. 

For the Kerr-case these polynomials are $\Delta \equiv 1, A_1 \equiv - \mathrm i (1-y)^2, A_2 \equiv \mathrm i y (1-y), C_1 \equiv P_-, C_2 \equiv P_2$ and thus the important zeros are 
\begin{eqnarray}
\left\{ 0, 1, \frac{\eta x_+ \pm \mathrm i \kappa}{\eta x_- \pm \mathrm i \kappa} \right\} \, .
\end{eqnarray}
It is easy to verify that for all of these zeros the absolute value is either $0$ or $\ge 1$.

For the five dimensional Myers-Perry black hole-case these polynomials are $\Delta \equiv \Delta_y, A_1 \equiv \mathrm i \xi (1-y)^2, A_2 \equiv \mathrm i \xi y (1-y)^2, C_1 \equiv P_-, C_2 \equiv P_+$. The zeros are 
\begin{eqnarray}
\left\{ 0, 1, \frac{x_+}{x_-}, \frac{2 x_+}{x_+ + x_-}, \frac{x_+ + x_-}{2 x_-}, \text{the three zeros of } P_\pm \right\} \, . \nonumber \\
\end{eqnarray}
The first five zeros have an absolute value of either $0$ or $\ge 1$. So we need to keep track of the zeros of $P_\pm$ for the respective second order equation of $\psi_{1, 2}$ to make sure that the Frobenius Ansatz works in these cases.

We reduce the recurrence relation for the coefficients resulting from substituting the Ansatz (\ref{eqn:frobenius_ansatz}) into the second order differential equation down to a recurrence relation of order 2 using Gaussian elimination. The recurrence relation of order 2 then leads to a continued fraction. The coefficients of the continued fraction depend on the frequency $\Omega$. The continued fraction converges when $\Omega$ is an eigenvalue, and hence we search in the complex $\Omega$-plane for these convergent values. 
We evaluate the continued fraction only to a finite depth. The rest can be approximated asymptotically using the Nollert improvement \cite{Leaver:1985ax, Nollert:1993zz}. 

This method is implemented into a C++ script using a class for the polynomial algebra and a class for the 
recurrence relations. The depth of the continued fraction evaluated is chosen to be of the order of 10000. The order of the Nollert improvement was up to terms of order $N^{-3/2}$. 
In practice, and in order to estimate the precision and get rid of spurious modes, we calculate the modes for both second order differential equations for $\psi_{1, 2}$. We will refer to them simply as equation 1 and equation 2 in the following section.
In the five dimensional case this can also be used to complement the modes when one of the equations cannot be used due to the zeros of $P_\pm$, 
provided surrounding modes in the parameter space spanned by $\eta$ and $\alpha$ match smoothly with these calculations.

As mentioned in the section on the Kerr case, the angular eigenvalue in four dimensions is calculated using the continued fraction equation (\ref{eqn:Kerr_angular_cfe}). 
This is also implemented in a C++ script.
The depth of the evaluated continued fraction was also 10000. But we did not use an asymptotic approximation in this case.

\section{Results}\label{sec:results}

In this section we present the results for the spectrum of quasinormal modes of the massive Dirac field. In particular, we will focus on the ground state and the first excitation. We will also discuss the behaviour of the angular eigenvalue with the different parameters of the field and the black hole. Let us start with the case of the Kerr background.

\subsection{The Dirac spectrum of the Kerr black hole}

Making use of the method described before we calculate in a systematic way the spectrum of quasinormal modes for different values of the parameters $\eta$ and $\alpha$. In the following figures we will only show the quasinormal modes with $\Re(\Omega) > 0$, because the modes with $\Re(\Omega) < 0$ follow by the transformation $(\alpha \lambda, \epsilon_\kappa; \Re(\Omega), \Re(\kappa)) \mapsto (-\alpha \lambda, -\epsilon_\kappa; -\Re(\Omega), -\Re(\kappa))$. This was noted as a symmetry of the radial equation in \cite{Konoplya:2017tvu}. 
We have also crosschecked our results with the ones presented in \cite{Konoplya:2017tvu}.
Also in these figures we always show the calculated quasinormal modes of both second order equations in one plot. In most of the cases the difference is very small, and both curves overlap.

We will start discussing the properties of the ground mode.

\begin{figure*}
\includegraphics[width=0.95\linewidth]{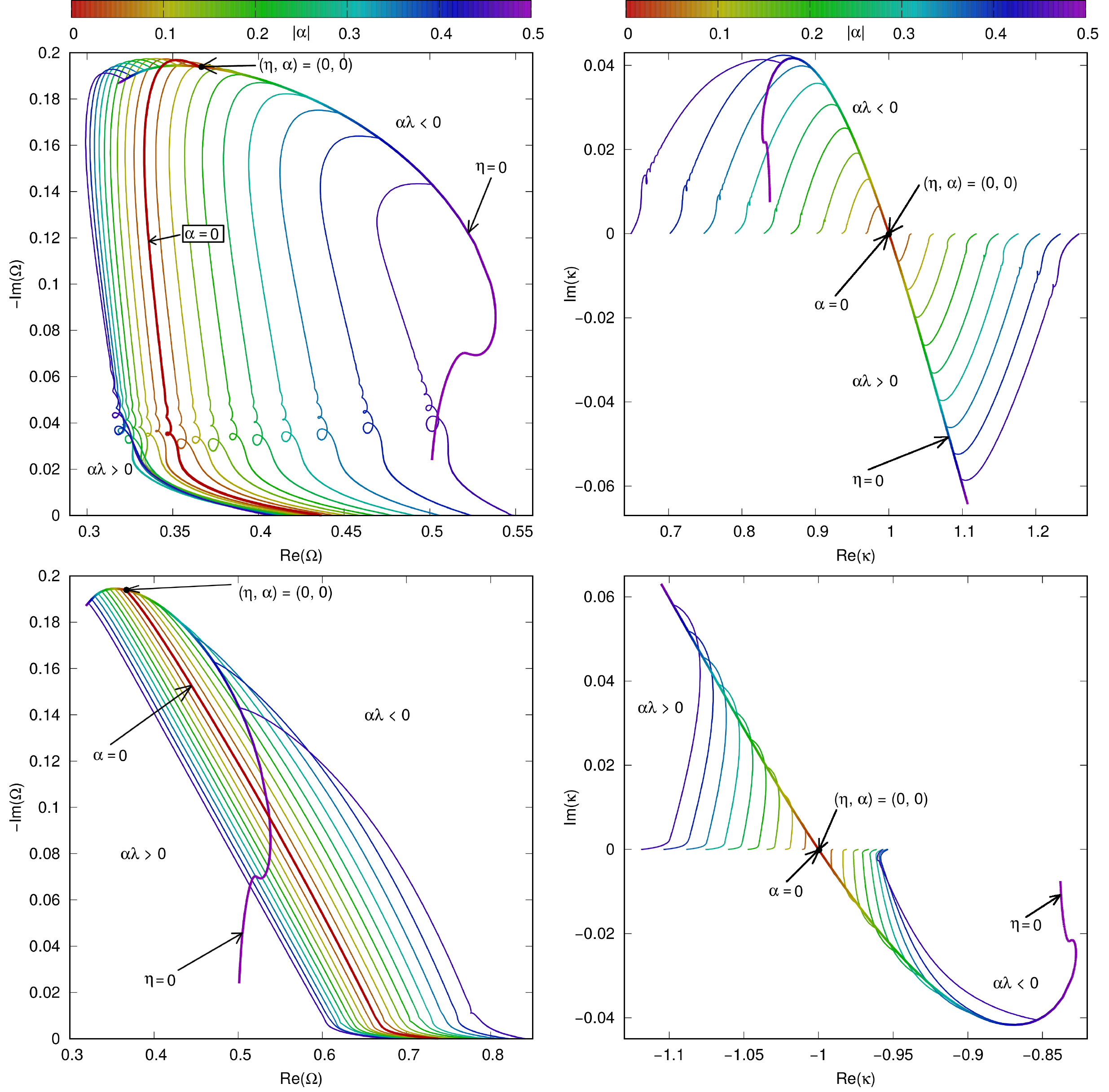}
\caption{\label{fig:d=4_n=0} The negative imaginary part $- \Im(\Omega)$ of the quasinormal modes over the real part $\Re(\Omega)$ (left column) and the imaginary part of the angular eigenvalue $\Im(\kappa)$ over the real part of the angular eigenvalue $\Re(\kappa)$ (right column) in the four dimensional Kerr metric. Displayed are the ground modes for $\kappa(\alpha=0) = +1$ (upper-left plot) and $\kappa(\alpha=0) = -1$ (lower-left plot) and the corresponding values of the angular eigenvalue $\kappa$ (right column). Different values of $\alpha$ are color-coded from red $(\alpha = 0)$ up to purple $(\alpha = 0.5)$. With thicker lines marked are the modes for $\alpha = 0$ and the massless modes $\eta = 0$. Indicated with a black point is the massless mode for the static limit. The magnetic quantum number is $|\lambda| = 0.5$. Branching off from the massless $\eta = 0$ modes are modes with increasing $\eta$ for fixed $\alpha$ with a difference of $\Delta \alpha = 0.05$ between two neighboring ones.}
\end{figure*}
\begin{figure*}
\includegraphics[trim={0 20 0 40}, clip, width=0.5\linewidth,height=0.22\linewidth]{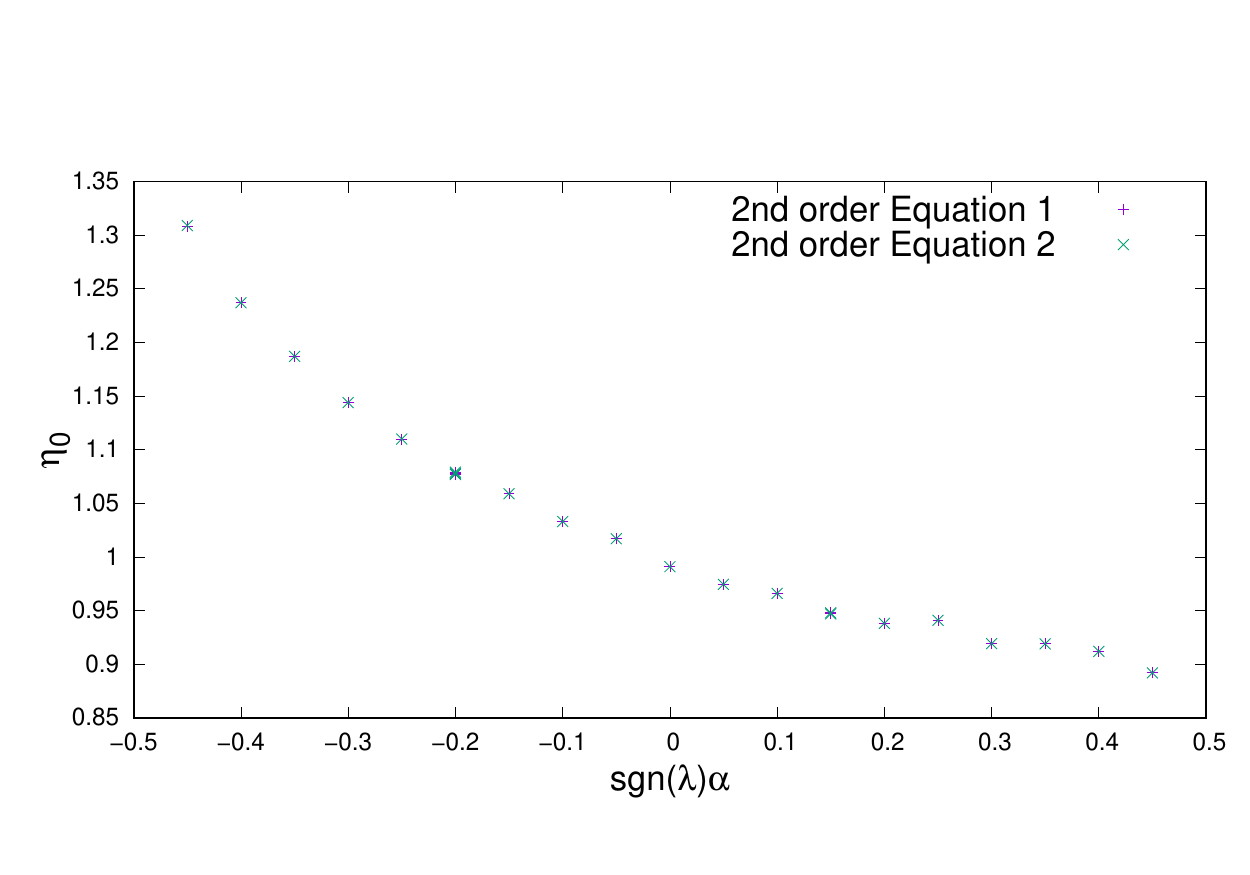}
\caption{\label{fig:eta0} The approximate critical mass value $\eta_0$ over $\alpha$ for the quasiresonant ground modes of the $\kappa(\alpha = 0) = +1$ branch. Shown are the mass values for which $|\Im(\Omega)| \le 10^{-5}$. 
}
\end{figure*}
\begin{figure*}
\includegraphics[width=0.95\linewidth]{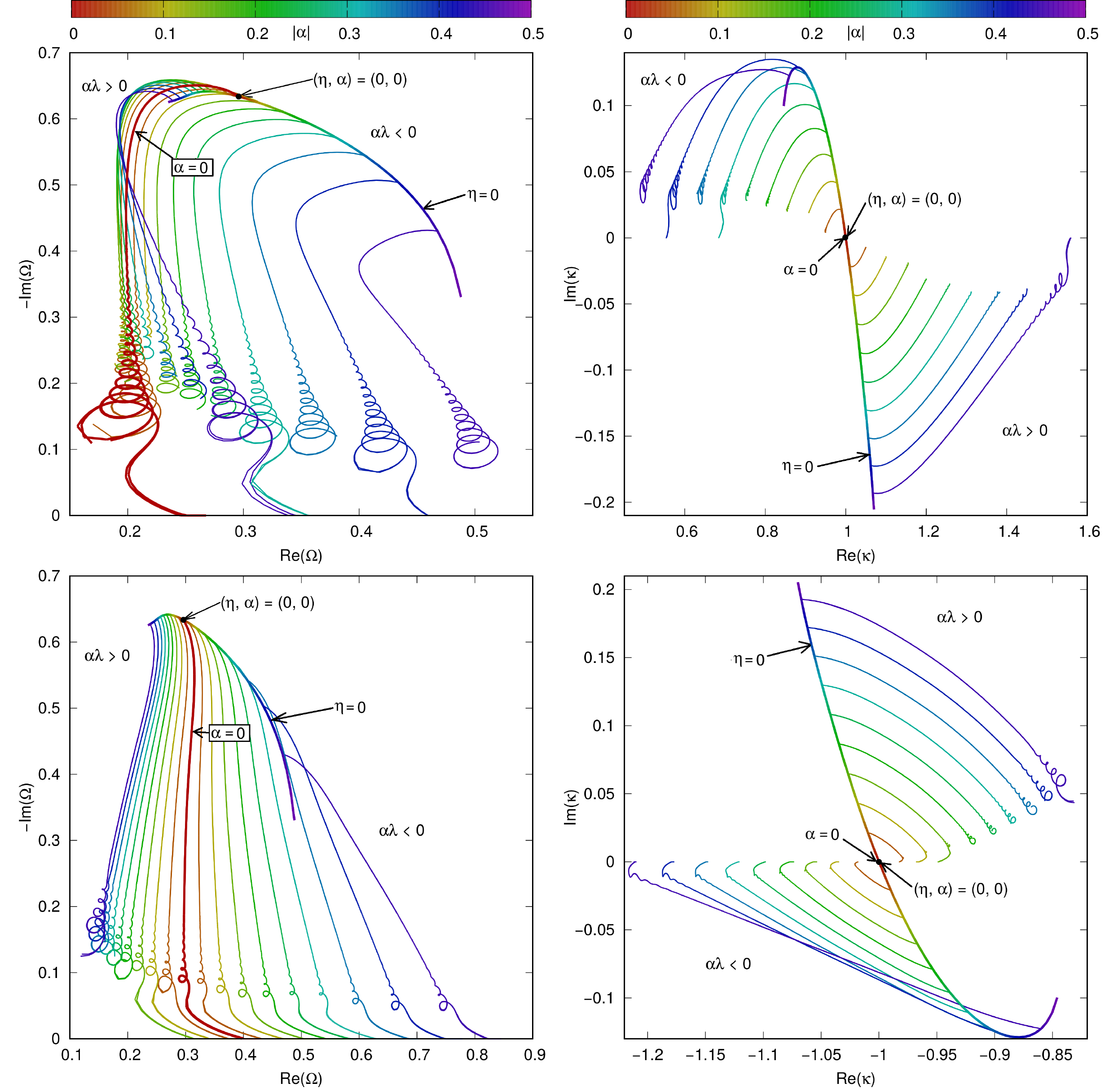}
\caption{\label{fig:d=4_n=1} Similar to Figure \ref{fig:d=4_n=0}, but for the first excited mode. The left column shows the modes, while the right column shows the eigenvalue $\kappa$. The upper row is for  $\kappa(\alpha=0) = +1$ and the lower row for $\kappa(\alpha=0) = -1$.
}
\end{figure*}

\begin{itemize}
\item \textbf{Ground mode}
\end{itemize}

Shown in figure \ref{fig:d=4_n=0} are the calculated quasinormal modes $\Omega$ (left column) and the angular eigenvalue $\kappa$ (right column) for the ground state with magnetic quantum number $|\lambda| = 0.5$. 
The upper row corresponds to the modes with $\kappa(\alpha = 0) = +1$, and the lower row to the modes with $\kappa(\alpha = 0) = -1$. 
With colors, we mark different values of the angular parameter $\alpha$, with red being $\alpha = 0$ and purple  $\alpha = 0.5$. Marked with thicker lines are the modes for $\alpha = 0$ (thick red curve) and the massless modes $\eta = 0$ (colored thick curve). The massless mode is explicitly marked in the static case $(\eta = 0, \alpha = 0)$ with a black dot. 
From the curve with $\eta = 0$ we plot branches of modes with fixed values of $\alpha$ and increasing values of $\eta$ with a difference of $\Delta \alpha = 0.5$ between two neighboring ones. 
Note the sign of $\alpha \lambda$ splits the modes into two, to the left or to the right of the $\eta=\alpha=0$ case. We mark each family with a label indicating the sign of $\alpha \lambda$.

Comparing the upper-left plot with the spectrum with $\kappa(\alpha = 0) = +1$ with the lower-left plot with $\kappa(\alpha = 0) = -1$ we can see that
both sets of modes differ with regard to a change in $\eta$. Only when $\eta = 0$ the modes are equal (the thick colored curve is actually the same in both figures). The overall effect is however similar in both cases: an increase in the mass $\eta$ generally leads to a reduction of the absolute value of the imaginary part of the frequency $|\Im (\Omega)|$. The behaviour of the modes with fixed values of $\alpha$ follows the general behaviour of the $\alpha = 0$ modes when we increase the value of $\eta$.

While for $\alpha=0$ it is the sign of $\kappa$ what determines the behaviour of the modes for a given value of $\eta$, in the $\eta=0$ case it is the sign of $\lambda$ that determines the behaviour for a given value of $\alpha$.
We will see that this property of the spinor field will be important in five dimensions and leads to a breakdown of the mapping for $\Re(\Omega) \mapsto -\Re(\Omega)$ in the case of $\alpha \neq 0 \neq \eta$. %

 For large mass values both the imaginary part $\Im (\Omega)$ and the real part $\Re (\Omega)$ develop oscillations with regard to the mass $\eta$. This feature is more pronounced in the case of $\kappa(\alpha = 0) = +1$ (upper row) than in the case of $\kappa(\alpha = 0)=-1$ (lower row). These oscillations in the imaginary and real part combined with a drift in the complex plane lead to a looping behaviour found previously for scalar, vector and tensor quasinormal modes \cite{Glampedakis:2003dn, Onozawa:1996ux, Andersson:1996xw, Cook:2016fge, Cook:2014cta}. Although not shown in figure \ref{fig:d=4_n=0}, this looping behaviour is also present when varying $\alpha$ if one fixes $\eta$ with a large enough value.

As we commented previously, increasing $\eta$ decreases $|\Im(\Omega)|$, but the $\Re(\Omega)$  does not change so quickly. However, when the modes have a small enough $|\Im(\Omega)|$ for large values of $\eta$, the real part of the frequency $\Re(\Omega)$ increases a bit faster. The curves then approach the real axis. However, before the value $\Im(\Omega) = 0$ can be reached for a critical value of $\eta$, the modes disappear. Nontheless, the value of $\Im(\Omega) = 0$ can be made (in principle) as small as desired close enough to the critical value of $\eta$. This is the same quasiresonance behaviour found for other massive fields \cite{Konoplya:2011qq} and also already noted for the massive spin half field in Tangherlini \cite{Blazquez-Salcedo:2017bld} and in Kerr-Newman-AdS \cite{Konoplya:2017tvu}. We show the behaviour of these critical mass values $\eta_0$ depending on $\alpha$ for the ground mode of $\kappa(\alpha = 0) = +1$ in figure \ref{fig:eta0}. Here we display the mass values $\eta_0$ for which $|\Im(\Omega)|\le 10^{-5}$. 
We can observe that for $\mathrm{sgn}(\lambda) \alpha> \mathrm{sgn}(\lambda) \alpha^\prime$ we have $\eta_0(\alpha) < \eta_0(\alpha^\prime)$. Although not shown, this behaviour of the critical mass is similar for $\kappa(\alpha = 0) = -1$.

Going back to figure \ref{fig:d=4_n=0}, let us focus now on the angular eigenvalue $\kappa$ (displayed in the right column plots). The upper-right plot is for $\kappa ( \alpha = 0) = +1$ and the lower-right one is for $\kappa(\alpha =0) = -1$. This eigenvalue possesses features very similar to what is found for the modes. One can observe that the behaviour with regard to $\alpha$ depends on the sign of $\alpha \lambda$. The oscillations from the quasinormal frequencies leading to a looping behaviour can also be found for the angular eigenvalue for large enough $\eta$. The quasiresonances with small $|\Im(\Omega)|$ also have small $|\Im(\kappa)|$. In particular, this happens because for real $\Omega$ the recurrence relation (\ref{eqn:angular_recc}) does not contain a complex part except for possibly $\kappa$, and thus $\kappa$ must be also real. The real part of $\kappa$ also changes more rapidly when the real axis is approached, as it happens in the modes. The quasiresonant modes with $\alpha \lambda \Re(\kappa) \ge 0$ fulfill $|\Re(\kappa^{(\alpha)}_\text{qr})| \ge |\Re(\kappa(\alpha = 0)|$ with equality for $\alpha = 0$, where "qr" denotes quasiresonance. Accordingly we also have $\alpha \lambda \Re(\kappa) \le 0$ fulfilling $|\Re(\kappa^{(\alpha)}_\text{qr})| \le |\Re(\kappa(\alpha = 0)|$ with equality for $\alpha = 0$.

Finally, let us comment here that the eigenvalue $\kappa$ possesses the symmetry in the massless case $\kappa(\alpha, \eta=0; \lambda, \epsilon_\kappa) = - \kappa(\alpha, \eta=0; \lambda, -\epsilon_\kappa)$, with the quasinormal modes mapped accordingly. However, as long as the field is massive this symmetry is broken.

\begin{itemize}
\item \textbf{First excitation}
\end{itemize}

In figure \ref{fig:d=4_n=1}, we show the results for the first excited state. The conventions we use here are similar to the ones introduced in figure \ref{fig:d=4_n=0}.

The general behaviour for the quasinormal modes $\Omega$ and for the angular eigenvalue $\kappa$ connected to the eigenvalue $\kappa(\alpha=0) = +1$ is the same as in the ground mode case. However the loops are more pronounced in the first excited mode. Quasiresonances also appear here, for certain values of the mass. However, the numerics for large values of $\eta$ are not so precise as in the gound mode. We show the modes either until they approach the quasiresonances or the numerics begin to diverge (the difference between the quasinormal modes stemming from the two second order equations become greater than about 0.01 in absolute value, as it can be appreciated in the figure). 

A different behaviour from the one of the ground mode can be observed when the angular eigenvalue $\kappa$ is connected to $\kappa(\alpha=0)=-1$. The general behaviour for the dependence on $\eta$ for $\alpha \neq 0$ is flipped with regard to the sign of $\alpha \lambda$, meaning that for $\alpha \lambda \Re(\kappa) \ge 0$ fulfilling $|\Re(\kappa^{(\alpha)}_\text{qr})| \le |\Re(\kappa(\alpha = 0)|$ with equality for $\alpha = 0$. 

Let us note that the symmetry $\kappa(\alpha, \eta=0; \lambda, \epsilon_\kappa) = - \kappa(\alpha, \eta=0; \lambda, -\epsilon_\kappa)$ and the accompanying behaviour of $\kappa$ is also present for the first excited modes.

\begin{figure}
\includegraphics[trim={52 0 73 5}, clip, width=0.95\linewidth]{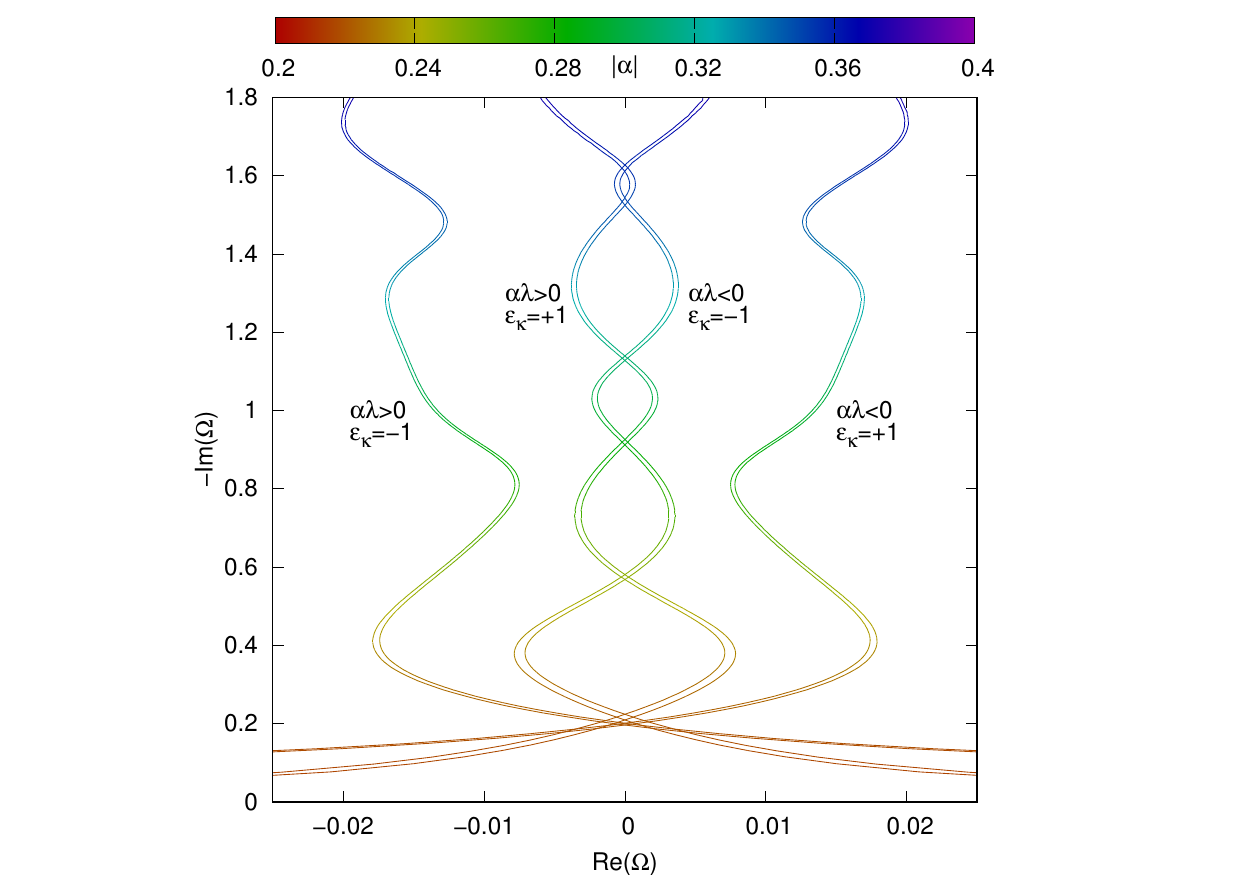}
\caption{\label{fig:d=4_high_n} The negative imaginary part $- \Im(\Omega)$ of the quasinormal modes over the real part $\Re(\Omega)$ in the four dimensional Kerr metric for high overtone number with fixed mass $\eta=1.93$. Different values of $\alpha$ are colorcoded from red $(\alpha = 0.2)$ up to purple $(\alpha = 0.4)$. As with the other figures we show the modes stemming from both second order differential equations. Here the small difference between the results of both second order equations can be appreciated, and it provides an estimation of the error of the modes. 
}
\end{figure}

\begin{itemize}
\item \textbf{High overtone number}
\end{itemize}

The very high overtone number modes develop parts which cross the imaginary axis in the complex plane and thus it is possible to have modes with $\Re(\Omega) = 0$ (pure pulses). To demonstrate this we show in figure \ref{fig:d=4_high_n} modes stemming from $\eta = 1.93, |\lambda|=0.5$ and varying $\alpha$. We show two sets of exemplary modes, where the two modes of each set are connected by the symmetry $(\alpha \lambda, \epsilon_\kappa; \Re(\Omega), \Re(\kappa)) \mapsto (-\alpha \lambda, -\epsilon_\kappa; -\Re(\Omega), -\Re(\kappa))$. The shown behaviour is quite generic for these high overtone number quasinormal modes close to the imaginary axis. Because we cannot trace these modes back to their $\eta = 0$ and $\alpha = 0$ origins (the numerics break down at a certain point, but normally $\Re(\Omega)$ is small and $|\Im(\Omega)|$ is rather large for the last modes where we can trust the numerics), we do not know to which overtone number the shown modes belong. The high overtone number modes are dense around the axis and thus 
we cannot be sure what are the exact overtone numbers. 
These modes with vanishing real part behave similarly to tensor perturbations \cite{Cook:2016fge}.

\begin{figure}
\includegraphics[trim={52 0 71 5}, clip, width=0.95\linewidth]{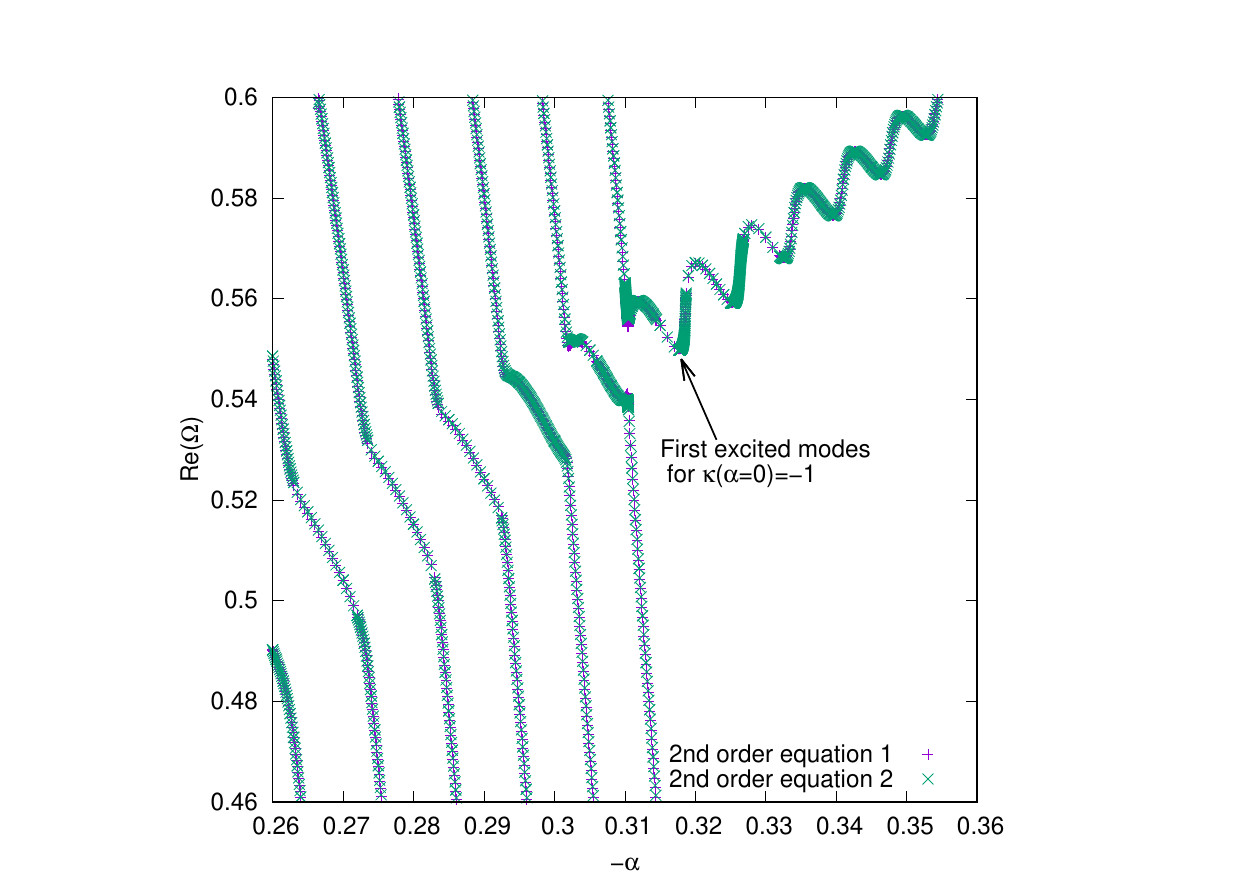}
\caption{\label{fig:d=4_high_n_window} The real part of the frequency $\Re(\Omega)$ as a function of $-\alpha$ for some quasinormal modes in the four dimensional Kerr metric. The mass value is fixed to be $\eta=2$. We mark with an arrow the branch of modes which can be continuously connected with the first excited mode for $(\alpha = 0, \eta = 0)$. The rest of branches possess higher overtone numbers.}
\end{figure}

These modes are also very dense near the real axis in the complex frequency plane. To demonstrate this we show in figure \ref{fig:d=4_high_n_window} for a fixed mass of $\eta=2$ and $\lambda = 0.5$ the real part of the frequency $\Re(\Omega)$ over the negative angular parameter $-\alpha$. In the upper right part there are the modes that are continuously connected to the first excited mode for $\alpha=0$ and $\eta=0$. As we decrease $|\alpha|$ the real part of the frequency
$\Re(\Omega)$ possesses an oscillatory behaviour. Eventually when $|\alpha|$ is small enough, $\Re(\Omega)$ rises quickly, while a different branch of modes with lower $\Re(\Omega)$ and higher overtone number approaches from below. This behaviour then repeats as we decrease $|\alpha|$, as it can be seen in figure \ref{fig:d=4_high_n_window}. 
The different branches of modes displayed here are not connected.
The modes with increasing $\Re(\Omega)$ eventually approach the critical value for which the quasiresonance appears, and the mode disappears as discussed previously.
The modes with small values of $\Re(\Omega)$ actually approach the imaginary axis as $|\Im(\Omega)|$ increases, similarly to the behaviour shown in figure \ref{fig:d=4_high_n}.

To summarize this section, we have seen that the Dirac field in the Kerr background possesses quasiresonant modes for critical values of the fermionic mass $\eta$, that depend on the value of $\alpha$ and the other quantum numbers of the field. This seems to happen for arbitrary overtone number, where the spectrum also possesses pure pulses. The introduction of rotation breaks some symmetries with the angular quantum numbers that exist only for the static case. Let us now continue with the quasinormal modes of the Dirac field in the background of the 5d Myers-Perry black hole and compare it to the Kerr case.

\subsection{The Dirac spectrum of the 5d Myers-Perry black hole}

\begin{figure*}
\includegraphics[width=0.95\linewidth]{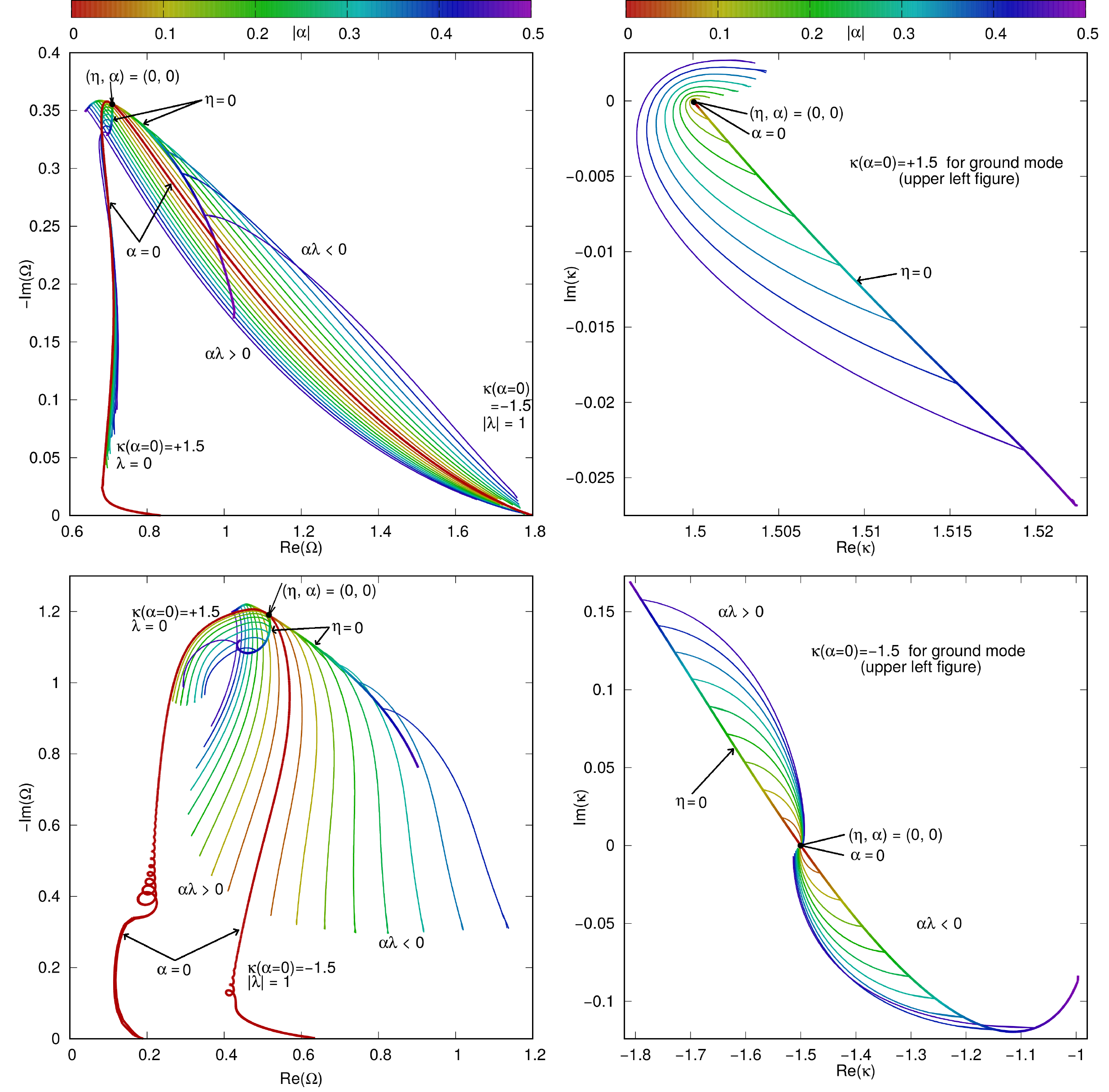}
\caption{\label{fig:d=5} The negative imaginary part $- \Im(\Omega)$ of the quasinormal modes over the real part $\Re(\Omega)$ and the imaginary part of the angular eigenvalue $\Im(\kappa)$ over the real part of the angular eigenvalue $\Re(\kappa)$ in the five dimensional Myers-Perry metric with equal angular momenta. Displayed are the ground modes (upper-left plot) and the first excited modes (lower-left plot). Shown in the two right plots are the angular eigenvalues $\kappa$ for the ground mode. Different values of $\alpha$ are colorcoded from red $(\alpha = 0)$ up to purple $(\alpha = 0.5)$. With thicker lines marked are the modes for $\alpha = 0$ and the massless modes $\eta = 0$. Marked with an arrow is the massless mode for the static limit. In the static limit the angular eigenvalue of the left branch is $\kappa = +1.5$ and of the right branch $\kappa = -1.5$. Due to $\lambda = 0$ in the left branch the modes do not depend on the sign of $\alpha$. Branching off from the massless $\eta = 0$ modes are modes with increasing $\eta$ for fixed $\alpha$ with a difference of $\Delta \alpha = 0.05$ between two neighboring ones.}
\end{figure*}

In this section we will present a similar analysis of the spectrum of the Dirac field in the background of the 5d Myers-Perry black hole with equal angular momenta. The spectrum turns out to be richer than in the Kerr case.

In figure \ref{fig:d=5} (top-left), we present the ground quasinormal modes, while in the (bottom-left) we show the first excitation. We only include positive values of the real part of the frequency. We also choose the lowest possible value of the angular eigenvalue, which means $|\kappa(\alpha = 0)|=1.5$. For $\kappa(\alpha = 0)=1.5$, $|\lambda|=1$, while for $\kappa(\alpha = 0)=-1.5$, $|\lambda|=0$.

Focusing first on the ground modes in figure \ref{fig:d=5} (top-left), the behaviour of these modes is qualitatively similar to the Kerr case: an increase in mass $\eta$ generally leads to a decrease in $|\Im(\Omega)|$ until eventually the mode vanishes in a quasiresonance. The behaviour of the modes with fixed $\alpha\neq 0$ is also qualitatively similar to the modes with $\alpha = 0$ with regards to an increase of $\eta$. Note that, due to the high order recurrence relation for $\alpha \neq 0$, the numerics were not convergent for all values of $\eta$ and we only show the parts where there was a sufficient agreement between the modes calculated from both second order equations. But it is likely that because of continuity in $\alpha$, the $\alpha \neq 0$ modes should follow the $\alpha = 0$ modes in behaviour even for large values of the mass $\eta$ and reach a quasiresonance. 

Observing the behaviour of various sets $\{\Omega (\alpha, \eta)\}$ under certain conditions leads to the same conclusions as in the Kerr case. For a fixed $|\kappa(\alpha = 0)|$ the sign of $\kappa(\alpha = 0)$ determines the behaviour of the modes for varying $\eta$ and fixed $\alpha$. 
For instance, the $\kappa(\alpha = 0)=-1.5$ branches fall quickly to a quasiresonance without changing that much the real part of the mode, while the $\kappa(\alpha = 0)=1.5$ reach the quasiresonance while the real part increases steadily.
For fixed $\eta = 0$ and varying $\alpha$ the value of $|\lambda|$ also changes the behaviour of the modes in the complex plane (this can be more easily understood if one considers large values of the angular eigenvalue $\kappa$, because there is a strong degeneracy in $\lambda$ for $\kappa(\alpha = 0)$ in this case).
%


In figure \ref{fig:d=5} (top-right) and (bottom-right) panels we show the angular eigenvalue $\kappa$ for $\kappa(\alpha = 0)=1.5$ and $\kappa(\alpha = 0)=-1.5$ respectively.


The angular eigenvalue $\kappa$ shares some similarities with the Kerr case, in the sense that increasing $\eta$ leads to small values $|\Im(\kappa)|$. But interestingly, for $\kappa(\alpha = 0) = -1.5$ the eigenvalues with $\alpha \neq 0$ seem to approach the original value of $-1.5$ for large mass values (as we approach the quasiresonance). The eigenvalues for $\kappa(\alpha = 0)= +1.5$ loop around the value $+1.5$ in the case of $\alpha \neq 0$ for varying $\eta$. Thus, there are certain values of the mass $\eta$ for which $\Im(\kappa) = 0$.

Since in this case we have analytical expressions for the angular eigenvalue, we can gain some understanding of this phenomenon by looking at the equations (\ref{eqn:MP_angular_EV_full}, \ref{eqn:MP_angular_EV_special_1} and \ref{eqn:MP_angular_EV_special_2}). In the case of $\kappa(\alpha = 0)=-1.5, |\lambda| = 1$ the angular eigenvalue is given by
\begin{eqnarray}\label{eqn:MP_angular_explicit_-}
\kappa_{|\lambda| = 1} = - 1.5 - \alpha \, \mathrm{sgn}(\lambda) \, (\Omega - \eta) \, .
\end{eqnarray}
Thus if $\Im(\Omega) \rightarrow 0$ also $\Im(\kappa) \rightarrow 0$. That the angular eigenvalue approaches $-1.5$ as $\eta$ grows shows, that $\Omega \sim \eta$ for large $\eta$. The angular eigenvalue with $\kappa(\alpha=0)=+1.5, \lambda = 0$ is given by
\begin{eqnarray}\label{eqn:MP_angular_explicit_+}
\kappa_{\lambda = 0} = -0.5 + \sqrt{\alpha^2 (\Omega - \eta)^2 + 4} \, .
\end{eqnarray}
Again, if $\Im(\Omega)=0$ the square root is real and thus $\Im(\kappa)=0$. The square root can also be real for $\Re(\Omega)=\eta$ and $|\Im(\Omega)| \le 2$. As one can see this is the case and we have for large enough $\eta$ modes with $\Re(\Omega) \le \eta$, which are gravitationally trapped around the black hole.

In the case of $\kappa(\alpha = 0) = -1.5$, one has $\lambda \neq 0$, and the dependence of the massless modes on $\alpha$ changes with the sign of $\alpha \lambda$. In the case of $\kappa(\alpha = 0) = +1.5$, one has $\lambda = 0$, and the massless modes only depend on $|\alpha|$ (they are independent of the sign of $\alpha$). This categorization of the dependence of $\alpha$ is independent of the sign of $\Re(\Omega)$. Thus, the mapping $\Re(\Omega) \mapsto -\Re(\Omega)$ must map $\kappa(\alpha=0)$ to the same value for these modes. The map is given by $(\alpha \lambda; \Re(\Omega), \Im(\kappa)) \mapsto (-\alpha \lambda; -\Re(\Omega), -\Im(\kappa))$. In the special case shown one can see this easily from equations (\ref{eqn:MP_angular_explicit_-}) and (\ref{eqn:MP_angular_explicit_+}). For $|\lambda|=1$, if one maps $\Re(\Omega) \mapsto - \Re(\Omega)$ one effectively maps $\alpha \, \mathrm{sgn}(\lambda) \, \Omega \mapsto -\alpha \, \mathrm{sgn}(\lambda) \, \overline \Omega$ and thus changes the sign of $\Im(\kappa)$ and the dependence on the sign of $\alpha \lambda$. For $\lambda = 0$ one effectively maps $\alpha^2 \Omega^2$ to $\alpha^2 \overline{\Omega}^2$ and thus again changes the sign of $\Im(\kappa)$. For $\alpha = 0$ the map between the positive and negative part of the spectrum is given by $(\Re(\Omega), \kappa)\mapsto (-\Re(\Omega), -\kappa)$. Observe that these maps cannot be continuously connected and thus there is no obvious map between $\Re(\Omega)$ and $-\Re(\Omega)$ for $\alpha \neq 0 \neq \eta$.

\begin{figure*}
\includegraphics[width=0.95\linewidth]{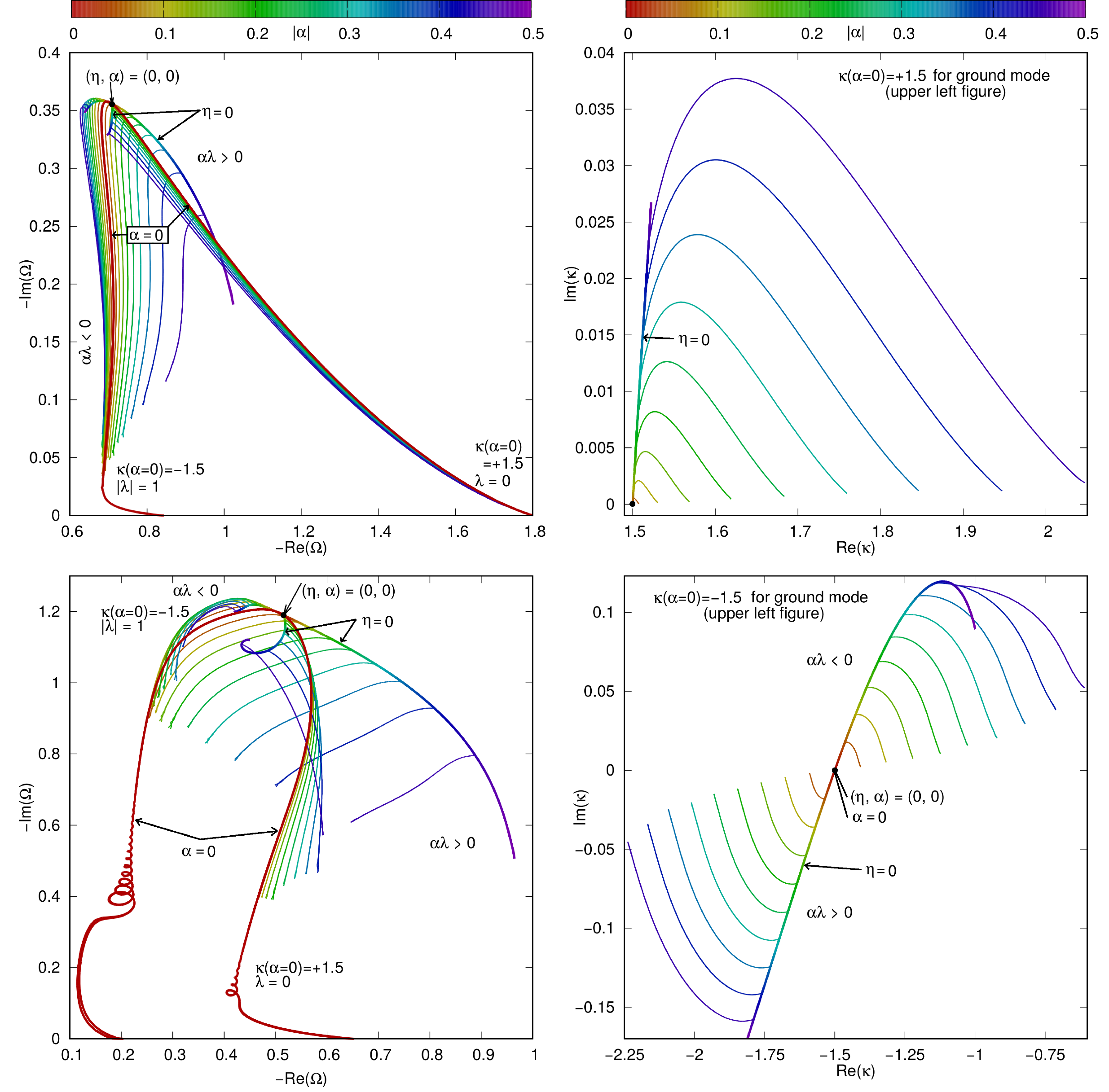}
\caption{\label{fig:d=5_negReO} Similar to Figure \ref{fig:d=5}, but for negative values of $\Re(\Omega)$ (note the inverted values of the axis in the plots of the left column).		
}
\end{figure*}

This behaviour is different from Kerr, where we have a symmetry. 
To show that this is not the case in Myers-Perry, we show in figure \ref{fig:d=5_negReO} the modes calculated with negative values of the real part of $\Omega$. We follow a similar structure to figure \ref{fig:d=5}. In the (top-left) and (top-bottom) we display the quasinormal modes for the ground mode and first excitation respectively. We also fix the lowest angular eigenvalue, which again means $|\kappa(\alpha = 0)|=1.5$. For $\kappa(\alpha = 0)=1.5$, $|\lambda|=1$, while for $\kappa(\alpha = 0)=-1.5$, $|\lambda|=0$.

Focusing on the ground state in figure \ref{fig:d=5_negReO} (upper-left), we can appreciate that, although the spectrum is different from the $\Re(\Omega) > 0$ case, the qualitative behavior is very similar.
The angular eigenvalue for the ground state is shown in \ref{fig:d=5_negReO} (upper-right) and (bottom-right) for $\kappa(\alpha = 0)=1.5$ and $\kappa(\alpha = 0)=-1.5$ respectively.

The angular eigenvalue also displays the same general behaviour as in the $\Re(\Omega) > 0$ case, although we do not observe a looping behaviour this time for the $\kappa(\alpha = 0)=+1.5$ modes around the static value, because there are no modes with $\Re(\Omega) = \eta$ and $|\Im(\Omega)|\le2$ in this case. Observe also that for $\Re(\Omega) < 0$ the $\kappa(\alpha = 0) = +1.5$ are more strongly affected by a change in $\eta$ for $\alpha \neq 0$, because of the larger $|\Re(\Omega)|$. The angular eigenvalues for $\kappa(\alpha = 0)=-1.5$ do not converge to the same value, but nevertheless $\Im(\kappa)$ tends to zero for increasing mass $\eta$, because $\Im(\Omega)$ also tends to zero.

Due to the non convergence of the numerics close to the quasiresonance for $\alpha \neq 0$ we cannot make a definite statement about the angular eigenvalue in these cases. Nevertheless, if we assume that the modes do not cross the $\Omega(\alpha =0)$ modes (again) before the quasiresonance, we can make the following argument for $\kappa_\text{qr}$ with regards to $\kappa(\alpha = 0)$ using equations (\ref{eqn:MP_angular_explicit_-}) and (\ref{eqn:MP_angular_explicit_+}) and the already present behaviour in the $\kappa$-plots. Here $\kappa_\text{qr}$ denotes the angular eigenvalue for the quasiresonance. For both sides of the complex $\Omega$-plane we note for $\kappa(\alpha = 0) = +1.5$ that close to the quasiresonance  $\Re(\kappa_\text{qr}) > 1.5 = \kappa(\alpha = 0)$. For $\kappa(\alpha = 0) = -1.5$ we have close to the quasiresonance if $\Re(\Omega) \alpha \lambda > 0 $, then $\kappa_\text{qr} \ge \kappa(\alpha=0) = -1.5$. If $\Re(\Omega) \alpha \lambda < 0 $, then $\kappa_\text{qr} \le \kappa(\alpha=0) = -1.5$. For both $\Re(\Omega)>0$ cases we have equality if close to the quasiresonance $\Omega = \eta$.


Going back to the first excited mode, we can compare the figure \ref{fig:d=5} (bottom-left) with the figure \ref{fig:d=5_negReO} (bottom-left).
One can observe in the first excited modes for  $\alpha = 0$ that, similarly to the Kerr case a looping behaviour develops for large values of the mass. This behaviour is then interrupted by a sudden change in $\Re(\Omega)$ before the quasiresonance is reached.

As with the Kerr case, close to the real axis a lot of high overtone number modes bunch up. However, due to the 
complicated numerics for the
$\alpha \neq 0$ modes, we were only able to reliably observe this for $\alpha = 0$, although we will not show these results here since they are similar to what was shown for Kerr in figure \ref{fig:d=4_high_n}. However, and contrary to what happens in the Kerr case, we were not able to find modes close to the imaginary axis stemming from high overtone number modes.


%

\section{conclusions}\label{sec:conclusions}

Using the methods known from the literature we decoupled the Dirac equation in the Kerr and five dimensional Myers-Perry spacetime with equal angular momenta into ordinary differential equations. We used methods from the literature to extract a recurrence relation for the angular part of the Kerr-case in order to determine the angular eigenvalue given a frequency. The angular equation in the five dimensional Myers-Perry case was solved analytically and a formula for the angular eigenvalue was obtained. The radial equations were used to construct two coupled second order differential equations with polynomial coefficients for both cases under study. These were then used to obtain recurrence relations for coefficients in a Frobenius-Fuchs Ansatz for the radial solutions using boundary conditions given by the physical situation of quasinormal modes. Using both equations simultaneously gave us an indication as to when the continued fraction method converged for given parameter values.

We then proceeded to calculate the quasinormal mode spectra for the ground state and first excited state in both cases. The numerical values for the Kerr case agreed with the previously calculated ones from the literature. Several phenomena were observed that are also present for other fields in rotating black hole spacetimes. For instance, we find quasiresonances for finite values of field masses, a looping like behaviour of the modes for large mass values in both varying mass and angular momenta, and a bunching up of high overtone number modes near the axes for large values of the given parameters. We were able to calculate modes in the Kerr case with vanishing real part (pure pulses). 

In the Myers-Perry case we observed that there is a breakdown of the symmetry between the positive and negative real part of the frequency spectra for non-vanishing field mass and angular momenta. This behaviour can be understood from the fact that for vanishing angular momentum the connection includes a change in sign for the angular eigenvalue in the modes. But for vanishing mass the behaviour is determined by the magnetic quantum numbers and thus the modes must keep the real part of the angular eigenvalue the same. These cases cannot be connected continuously as maps and thus the symmetry breaks for massive Dirac fields in the rotating black hole background.

Possible extensions of this work could be to consider the five dimensional case with unequal angular momenta, and to extend to more general spacetimes with additional parameters, for example a (NUT) charge and/or cosmological constant. Also of interest would be the study of how the behaviour changes in dimensions higher than five. Another interesting case would be to study the quasinormal modes for the extremal cases of the black holes and possibly extract analytical formulas for these, as these modes would set boundaries for the sets $\{\Omega(\alpha, \eta)\}$ in the complex plane.

\

\section{acknowledgements}\label{acknowledgments}
We would like to thank Jutta Kunz for valuable comments.
JLBS and CK would like to acknowledge support by the 
DFG Research Training Group 1620 {\sl Models of Gravity}.
JLBS would like to acknowledge support from the DFG project BL 1553 and the COST Action CA16104.

\bibliography{PaperV4.bbl}

\providecommand{\noopsort}[1]{}\providecommand{\singleletter}[1]{#1}%
\begin{thebibliography}{48}%
\makeatletter
\providecommand \@ifxundefined [1]{%
 \@ifx{#1\undefined}
}%
\providecommand \@ifnum [1]{%
 \ifnum #1\expandafter \@firstoftwo
 \else \expandafter \@secondoftwo
 \fi
}%
\providecommand \@ifx [1]{%
 \ifx #1\expandafter \@firstoftwo
 \else \expandafter \@secondoftwo
 \fi
}%
\providecommand \natexlab [1]{#1}%
\providecommand \enquote  [1]{``#1''}%
\providecommand \bibnamefont  [1]{#1}%
\providecommand \bibfnamefont [1]{#1}%
\providecommand \citenamefont [1]{#1}%
\providecommand \href@noop [0]{\@secondoftwo}%
\providecommand \href [0]{\begingroup \@sanitize@url \@href}%
\providecommand \@href[1]{\@@startlink{#1}\@@href}%
\providecommand \@@href[1]{\endgroup#1\@@endlink}%
\providecommand \@sanitize@url [0]{\catcode `\\12\catcode `\$12\catcode
  `\&12\catcode `\#12\catcode `\^12\catcode `\_12\catcode `\%12\relax}%
\providecommand \@@startlink[1]{}%
\providecommand \@@endlink[0]{}%
\providecommand \url  [0]{\begingroup\@sanitize@url \@url }%
\providecommand \@url [1]{\endgroup\@href {#1}{\urlprefix }}%
\providecommand \urlprefix  [0]{URL }%
\providecommand \Eprint [0]{\href }%
\providecommand \doibase [0]{http://dx.doi.org/}%
\providecommand \selectlanguage [0]{\@gobble}%
\providecommand \bibinfo  [0]{\@secondoftwo}%
\providecommand \bibfield  [0]{\@secondoftwo}%
\providecommand \translation [1]{[#1]}%
\providecommand \BibitemOpen [0]{}%
\providecommand \bibitemStop [0]{}%
\providecommand \bibitemNoStop [0]{.\EOS\space}%
\providecommand \EOS [0]{\spacefactor3000\relax}%
\providecommand \BibitemShut  [1]{\csname bibitem#1\endcsname}%
\let\auto@bib@innerbib\@empty
\bibitem [{\citenamefont {Horowitz}(2012)}]{Horowitz:2012nnc}%
  \BibitemOpen
  \bibinfo {editor} {\bibfnamefont {G.~T.}\ \bibnamefont {Horowitz}},\ ed.,\
  \href {http://www.cambridge.org/de/knowledge/isbn/item6633780} {\emph
  {\bibinfo {title} {{Black holes in higher dimensions}}}}\ (\bibinfo
  {publisher} {Cambridge Univ. Pr.},\ \bibinfo {address} {Cambridge, UK},\
  \bibinfo {year} {2012})\BibitemShut {NoStop}%
\bibitem [{\citenamefont {Emparan}\ and\ \citenamefont
  {Reall}(2008)}]{Emparan:2008eg}%
  \BibitemOpen
  \bibfield  {author} {\bibinfo {author} {\bibfnamefont {R.}~\bibnamefont
  {Emparan}}\ and\ \bibinfo {author} {\bibfnamefont {H.~S.}\ \bibnamefont
  {Reall}},\ }\href {\doibase 10.12942/lrr-2008-6} {\bibfield  {journal}
  {\bibinfo  {journal} {Living Rev. Rel.}\ }\textbf {\bibinfo {volume} {11}},\
  \bibinfo {pages} {6} (\bibinfo {year} {2008})},\ \Eprint
  {http://arxiv.org/abs/0801.3471} {arXiv:0801.3471 [hep-th]} \BibitemShut
  {NoStop}%
\bibitem [{\citenamefont {Konoplya}\ and\ \citenamefont
  {Zhidenko}(2011)}]{Konoplya:2011qq}%
  \BibitemOpen
  \bibfield  {author} {\bibinfo {author} {\bibfnamefont {R.~A.}\ \bibnamefont
  {Konoplya}}\ and\ \bibinfo {author} {\bibfnamefont {A.}~\bibnamefont
  {Zhidenko}},\ }\href {\doibase 10.1103/RevModPhys.83.793} {\bibfield
  {journal} {\bibinfo  {journal} {Rev. Mod. Phys.}\ }\textbf {\bibinfo {volume}
  {83}},\ \bibinfo {pages} {793} (\bibinfo {year} {2011})},\ \Eprint
  {http://arxiv.org/abs/1102.4014} {arXiv:1102.4014 [gr-qc]} \BibitemShut
  {NoStop}%
\bibitem [{\citenamefont {Bl\'azquez-Salcedo}\ \emph
  {et~al.}(2018)\citenamefont {Bl\'azquez-Salcedo}, \citenamefont {Doneva},
  \citenamefont {Kunz},\ and\ \citenamefont
  {Yazadjiev}}]{Blazquez-Salcedo:2018jnn}%
  \BibitemOpen
  \bibfield  {author} {\bibinfo {author} {\bibfnamefont {J.~L.}\ \bibnamefont
  {Bl\'azquez-Salcedo}}, \bibinfo {author} {\bibfnamefont {D.~D.}\ \bibnamefont
  {Doneva}}, \bibinfo {author} {\bibfnamefont {J.}~\bibnamefont {Kunz}}, \ and\
  \bibinfo {author} {\bibfnamefont {S.~S.}\ \bibnamefont {Yazadjiev}},\
  }\href@noop {} {\  (\bibinfo {year} {2018})},\ \Eprint
  {http://arxiv.org/abs/1805.05755} {arXiv:1805.05755 [gr-qc]} \BibitemShut
  {NoStop}%
\bibitem [{\citenamefont {Degollado}\ \emph {et~al.}(2018)\citenamefont
  {Degollado}, \citenamefont {Herdeiro},\ and\ \citenamefont
  {Radu}}]{Degollado:2018ypf}%
  \BibitemOpen
  \bibfield  {author} {\bibinfo {author} {\bibfnamefont {J.~C.}\ \bibnamefont
  {Degollado}}, \bibinfo {author} {\bibfnamefont {C.~A.~R.}\ \bibnamefont
  {Herdeiro}}, \ and\ \bibinfo {author} {\bibfnamefont {E.}~\bibnamefont
  {Radu}},\ }\href {\doibase 10.1016/j.physletb.2018.04.052} {\bibfield
  {journal} {\bibinfo  {journal} {Phys. Lett.}\ }\textbf {\bibinfo {volume}
  {B781}},\ \bibinfo {pages} {651} (\bibinfo {year} {2018})},\ \Eprint
  {http://arxiv.org/abs/1802.07266} {arXiv:1802.07266 [gr-qc]} \BibitemShut
  {NoStop}%
\bibitem [{\citenamefont {Bl\'azquez-Salcedo}\ and\ \citenamefont
  {Knoll}(2018{\natexlab{a}})}]{Blazquez-Salcedo:2017bld}%
  \BibitemOpen
  \bibfield  {author} {\bibinfo {author} {\bibfnamefont {J.~L.}\ \bibnamefont
  {Bl\'azquez-Salcedo}}\ and\ \bibinfo {author} {\bibfnamefont
  {C.}~\bibnamefont {Knoll}},\ }\href {\doibase 10.1103/PhysRevD.97.044020}
  {\bibfield  {journal} {\bibinfo  {journal} {Phys. Rev.}\ }\textbf {\bibinfo
  {volume} {D97}},\ \bibinfo {pages} {044020} (\bibinfo {year}
  {2018}{\natexlab{a}})},\ \Eprint {http://arxiv.org/abs/1709.07864}
  {arXiv:1709.07864 [gr-qc]} \BibitemShut {NoStop}%
\bibitem [{\citenamefont {Konoplya}\ and\ \citenamefont
  {Zhidenko}(2018)}]{Konoplya:2017tvu}%
  \BibitemOpen
  \bibfield  {author} {\bibinfo {author} {\bibfnamefont {R.~A.}\ \bibnamefont
  {Konoplya}}\ and\ \bibinfo {author} {\bibfnamefont {A.}~\bibnamefont
  {Zhidenko}},\ }\href {\doibase 10.1103/PhysRevD.97.084034} {\bibfield
  {journal} {\bibinfo  {journal} {Phys. Rev.}\ }\textbf {\bibinfo {volume}
  {D97}},\ \bibinfo {pages} {084034} (\bibinfo {year} {2018})},\ \Eprint
  {http://arxiv.org/abs/1712.06667} {arXiv:1712.06667 [gr-qc]} \BibitemShut
  {NoStop}%
\bibitem [{\citenamefont {Frolov}\ and\ \citenamefont
  {Stojkovic}(2003{\natexlab{a}})}]{Frolov:2003en}%
  \BibitemOpen
  \bibfield  {author} {\bibinfo {author} {\bibfnamefont {V.~P.}\ \bibnamefont
  {Frolov}}\ and\ \bibinfo {author} {\bibfnamefont {D.}~\bibnamefont
  {Stojkovic}},\ }\href {\doibase 10.1103/PhysRevD.68.064011} {\bibfield
  {journal} {\bibinfo  {journal} {Phys. Rev.}\ }\textbf {\bibinfo {volume}
  {D68}},\ \bibinfo {pages} {064011} (\bibinfo {year} {2003}{\natexlab{a}})},\
  \Eprint {http://arxiv.org/abs/gr-qc/0301016} {arXiv:gr-qc/0301016 [gr-qc]}
  \BibitemShut {NoStop}%
\bibitem [{\citenamefont {Frolov}\ and\ \citenamefont
  {Stojkovic}(2003{\natexlab{b}})}]{Frolov:2002xf}%
  \BibitemOpen
  \bibfield  {author} {\bibinfo {author} {\bibfnamefont {V.~P.}\ \bibnamefont
  {Frolov}}\ and\ \bibinfo {author} {\bibfnamefont {D.}~\bibnamefont
  {Stojkovic}},\ }\href {\doibase 10.1103/PhysRevD.67.084004} {\bibfield
  {journal} {\bibinfo  {journal} {Phys. Rev.}\ }\textbf {\bibinfo {volume}
  {D67}},\ \bibinfo {pages} {084004} (\bibinfo {year} {2003}{\natexlab{b}})},\
  \Eprint {http://arxiv.org/abs/gr-qc/0211055} {arXiv:gr-qc/0211055 [gr-qc]}
  \BibitemShut {NoStop}%
\bibitem [{\citenamefont {Chandrasekhar}(1984)}]{Chandrasekhar:1984siy}%
  \BibitemOpen
  \bibfield  {author} {\bibinfo {author} {\bibfnamefont {S.}~\bibnamefont
  {Chandrasekhar}},\ }\bibfield  {booktitle} {\emph {\bibinfo {booktitle}
  {{Proceedings, 10th International Conference on General Relativity and
  Gravitation: Padua, Italy, July 4-9, 1983}}},\ }\href {\doibase
  10.1007/978-94-009-6469-3_2} {\bibfield  {journal} {\bibinfo  {journal}
  {Fundam. Theor. Phys.}\ }\textbf {\bibinfo {volume} {9}},\ \bibinfo {pages}
  {5} (\bibinfo {year} {1984})}\BibitemShut {NoStop}%
\bibitem [{\citenamefont {Wu}(2008)}]{Wu:2008df}%
  \BibitemOpen
  \bibfield  {author} {\bibinfo {author} {\bibfnamefont {S.-Q.}\ \bibnamefont
  {Wu}},\ }\href {\doibase 10.1103/PhysRevD.78.064052} {\bibfield  {journal}
  {\bibinfo  {journal} {Phys. Rev.}\ }\textbf {\bibinfo {volume} {D78}},\
  \bibinfo {pages} {064052} (\bibinfo {year} {2008})},\ \Eprint
  {http://arxiv.org/abs/0807.2114} {arXiv:0807.2114 [hep-th]} \BibitemShut
  {NoStop}%
\bibitem [{\citenamefont {Oota}\ and\ \citenamefont
  {Yasui}(2008)}]{Oota:2007vx}%
  \BibitemOpen
  \bibfield  {author} {\bibinfo {author} {\bibfnamefont {T.}~\bibnamefont
  {Oota}}\ and\ \bibinfo {author} {\bibfnamefont {Y.}~\bibnamefont {Yasui}},\
  }\href {\doibase 10.1016/j.physletb.2007.11.057} {\bibfield  {journal}
  {\bibinfo  {journal} {Phys. Lett.}\ }\textbf {\bibinfo {volume} {B659}},\
  \bibinfo {pages} {688} (\bibinfo {year} {2008})},\ \Eprint
  {http://arxiv.org/abs/0711.0078} {arXiv:0711.0078 [hep-th]} \BibitemShut
  {NoStop}%
\bibitem [{\citenamefont {Cariglia}\ \emph {et~al.}(2011)\citenamefont
  {Cariglia}, \citenamefont {Krtous},\ and\ \citenamefont
  {Kubiznak}}]{Cariglia:2011qb}%
  \BibitemOpen
  \bibfield  {author} {\bibinfo {author} {\bibfnamefont {M.}~\bibnamefont
  {Cariglia}}, \bibinfo {author} {\bibfnamefont {P.}~\bibnamefont {Krtous}}, \
  and\ \bibinfo {author} {\bibfnamefont {D.}~\bibnamefont {Kubiznak}},\ }\href
  {\doibase 10.1103/PhysRevD.84.024008} {\bibfield  {journal} {\bibinfo
  {journal} {Phys. Rev.}\ }\textbf {\bibinfo {volume} {D84}},\ \bibinfo {pages}
  {024008} (\bibinfo {year} {2011})},\ \Eprint {http://arxiv.org/abs/1104.4123}
  {arXiv:1104.4123 [hep-th]} \BibitemShut {NoStop}%
\bibitem [{\citenamefont {Lunin}(2017)}]{Lunin:2017drx}%
  \BibitemOpen
  \bibfield  {author} {\bibinfo {author} {\bibfnamefont {O.}~\bibnamefont
  {Lunin}},\ }\href {\doibase 10.1007/JHEP12(2017)138} {\bibfield  {journal}
  {\bibinfo  {journal} {JHEP}\ }\textbf {\bibinfo {volume} {12}},\ \bibinfo
  {pages} {138} (\bibinfo {year} {2017})},\ \Eprint
  {http://arxiv.org/abs/1708.06766} {arXiv:1708.06766 [hep-th]} \BibitemShut
  {NoStop}%
\bibitem [{\citenamefont {Chervonyi}\ and\ \citenamefont
  {Lunin}(2015)}]{Chervonyi:2015ima}%
  \BibitemOpen
  \bibfield  {author} {\bibinfo {author} {\bibfnamefont {Y.}~\bibnamefont
  {Chervonyi}}\ and\ \bibinfo {author} {\bibfnamefont {O.}~\bibnamefont
  {Lunin}},\ }\href {\doibase 10.1007/JHEP09(2015)182} {\bibfield  {journal}
  {\bibinfo  {journal} {JHEP}\ }\textbf {\bibinfo {volume} {09}},\ \bibinfo
  {pages} {182} (\bibinfo {year} {2015})},\ \Eprint
  {http://arxiv.org/abs/1505.06154} {arXiv:1505.06154 [hep-th]} \BibitemShut
  {NoStop}%
\bibitem [{\citenamefont {Frolov}\ \emph {et~al.}(2017)\citenamefont {Frolov},
  \citenamefont {Krtous},\ and\ \citenamefont {Kubiznak}}]{Frolov:2017kze}%
  \BibitemOpen
  \bibfield  {author} {\bibinfo {author} {\bibfnamefont {V.}~\bibnamefont
  {Frolov}}, \bibinfo {author} {\bibfnamefont {P.}~\bibnamefont {Krtous}}, \
  and\ \bibinfo {author} {\bibfnamefont {D.}~\bibnamefont {Kubiznak}},\ }\href
  {\doibase 10.1007/s41114-017-0009-9} {\bibfield  {journal} {\bibinfo
  {journal} {Living Rev. Rel.}\ }\textbf {\bibinfo {volume} {20}},\ \bibinfo
  {pages} {6} (\bibinfo {year} {2017})},\ \Eprint
  {http://arxiv.org/abs/1705.05482} {arXiv:1705.05482 [gr-qc]} \BibitemShut
  {NoStop}%
\bibitem [{\citenamefont {Onozawa}(1997)}]{Onozawa:1996ux}%
  \BibitemOpen
  \bibfield  {author} {\bibinfo {author} {\bibfnamefont {H.}~\bibnamefont
  {Onozawa}},\ }\href {\doibase 10.1103/PhysRevD.55.3593} {\bibfield  {journal}
  {\bibinfo  {journal} {Phys. Rev.}\ }\textbf {\bibinfo {volume} {D55}},\
  \bibinfo {pages} {3593} (\bibinfo {year} {1997})},\ \Eprint
  {http://arxiv.org/abs/gr-qc/9610048} {arXiv:gr-qc/9610048 [gr-qc]}
  \BibitemShut {NoStop}%
\bibitem [{\citenamefont {Cook}\ and\ \citenamefont
  {Zalutskiy}(2014)}]{Cook:2014cta}%
  \BibitemOpen
  \bibfield  {author} {\bibinfo {author} {\bibfnamefont {G.~B.}\ \bibnamefont
  {Cook}}\ and\ \bibinfo {author} {\bibfnamefont {M.}~\bibnamefont
  {Zalutskiy}},\ }\href {\doibase 10.1103/PhysRevD.90.124021} {\bibfield
  {journal} {\bibinfo  {journal} {Phys. Rev.}\ }\textbf {\bibinfo {volume}
  {D90}},\ \bibinfo {pages} {124021} (\bibinfo {year} {2014})},\ \Eprint
  {http://arxiv.org/abs/1410.7698} {arXiv:1410.7698 [gr-qc]} \BibitemShut
  {NoStop}%
\bibitem [{\citenamefont {Wang}\ and\ \citenamefont
  {Herdeiro}(2016)}]{Wang:2015fgp}%
  \BibitemOpen
  \bibfield  {author} {\bibinfo {author} {\bibfnamefont {M.}~\bibnamefont
  {Wang}}\ and\ \bibinfo {author} {\bibfnamefont {C.}~\bibnamefont
  {Herdeiro}},\ }\href {\doibase 10.1103/PhysRevD.93.064066} {\bibfield
  {journal} {\bibinfo  {journal} {Phys. Rev.}\ }\textbf {\bibinfo {volume}
  {D93}},\ \bibinfo {pages} {064066} (\bibinfo {year} {2016})},\ \Eprint
  {http://arxiv.org/abs/1512.02262} {arXiv:1512.02262 [gr-qc]} \BibitemShut
  {NoStop}%
\bibitem [{\citenamefont {Glampedakis}\ and\ \citenamefont
  {Andersson}(2003)}]{Glampedakis:2003dn}%
  \BibitemOpen
  \bibfield  {author} {\bibinfo {author} {\bibfnamefont {K.}~\bibnamefont
  {Glampedakis}}\ and\ \bibinfo {author} {\bibfnamefont {N.}~\bibnamefont
  {Andersson}},\ }\href {\doibase 10.1088/0264-9381/20/15/312} {\bibfield
  {journal} {\bibinfo  {journal} {Class. Quant. Grav.}\ }\textbf {\bibinfo
  {volume} {20}},\ \bibinfo {pages} {3441} (\bibinfo {year} {2003})},\ \Eprint
  {http://arxiv.org/abs/gr-qc/0304030} {arXiv:gr-qc/0304030 [gr-qc]}
  \BibitemShut {NoStop}%
\bibitem [{\citenamefont {Morisawa}\ and\ \citenamefont
  {Ida}(2005)}]{Morisawa:2004fs}%
  \BibitemOpen
  \bibfield  {author} {\bibinfo {author} {\bibfnamefont {Y.}~\bibnamefont
  {Morisawa}}\ and\ \bibinfo {author} {\bibfnamefont {D.}~\bibnamefont {Ida}},\
  }\href {\doibase 10.1103/PhysRevD.71.044022} {\bibfield  {journal} {\bibinfo
  {journal} {Phys. Rev.}\ }\textbf {\bibinfo {volume} {D71}},\ \bibinfo {pages}
  {044022} (\bibinfo {year} {2005})},\ \Eprint
  {http://arxiv.org/abs/gr-qc/0412070} {arXiv:gr-qc/0412070 [gr-qc]}
  \BibitemShut {NoStop}%
\bibitem [{\citenamefont {Kunduri}\ \emph {et~al.}(2006)\citenamefont
  {Kunduri}, \citenamefont {Lucietti},\ and\ \citenamefont
  {Reall}}]{Kunduri:2006qa}%
  \BibitemOpen
  \bibfield  {author} {\bibinfo {author} {\bibfnamefont {H.~K.}\ \bibnamefont
  {Kunduri}}, \bibinfo {author} {\bibfnamefont {J.}~\bibnamefont {Lucietti}}, \
  and\ \bibinfo {author} {\bibfnamefont {H.~S.}\ \bibnamefont {Reall}},\ }\href
  {\doibase 10.1103/PhysRevD.74.084021} {\bibfield  {journal} {\bibinfo
  {journal} {Phys. Rev.}\ }\textbf {\bibinfo {volume} {D74}},\ \bibinfo {pages}
  {084021} (\bibinfo {year} {2006})},\ \Eprint
  {http://arxiv.org/abs/hep-th/0606076} {arXiv:hep-th/0606076 [hep-th]}
  \BibitemShut {NoStop}%
\bibitem [{\citenamefont {Kodama}\ \emph {et~al.}(2010)\citenamefont {Kodama},
  \citenamefont {Konoplya},\ and\ \citenamefont {Zhidenko}}]{Kodama:2009bf}%
  \BibitemOpen
  \bibfield  {author} {\bibinfo {author} {\bibfnamefont {H.}~\bibnamefont
  {Kodama}}, \bibinfo {author} {\bibfnamefont {R.~A.}\ \bibnamefont
  {Konoplya}}, \ and\ \bibinfo {author} {\bibfnamefont {A.}~\bibnamefont
  {Zhidenko}},\ }\href {\doibase 10.1103/PhysRevD.81.044007} {\bibfield
  {journal} {\bibinfo  {journal} {Phys. Rev.}\ }\textbf {\bibinfo {volume}
  {D81}},\ \bibinfo {pages} {044007} (\bibinfo {year} {2010})},\ \Eprint
  {http://arxiv.org/abs/0904.2154} {arXiv:0904.2154 [gr-qc]} \BibitemShut
  {NoStop}%
\bibitem [{\citenamefont {Kodama}\ \emph {et~al.}(2009)\citenamefont {Kodama},
  \citenamefont {Konoplya},\ and\ \citenamefont {Zhidenko}}]{Kodama:2009rq}%
  \BibitemOpen
  \bibfield  {author} {\bibinfo {author} {\bibfnamefont {H.}~\bibnamefont
  {Kodama}}, \bibinfo {author} {\bibfnamefont {R.~A.}\ \bibnamefont
  {Konoplya}}, \ and\ \bibinfo {author} {\bibfnamefont {A.}~\bibnamefont
  {Zhidenko}},\ }\href {\doibase 10.1103/PhysRevD.79.044003} {\bibfield
  {journal} {\bibinfo  {journal} {Phys. Rev.}\ }\textbf {\bibinfo {volume}
  {D79}},\ \bibinfo {pages} {044003} (\bibinfo {year} {2009})},\ \Eprint
  {http://arxiv.org/abs/0812.0445} {arXiv:0812.0445 [hep-th]} \BibitemShut
  {NoStop}%
\bibitem [{\citenamefont {Cho}\ \emph {et~al.}(2011)\citenamefont {Cho},
  \citenamefont {Doukas}, \citenamefont {Naylor},\ and\ \citenamefont
  {Cornell}}]{Cho:2011yp}%
  \BibitemOpen
  \bibfield  {author} {\bibinfo {author} {\bibfnamefont {H.~T.}\ \bibnamefont
  {Cho}}, \bibinfo {author} {\bibfnamefont {J.}~\bibnamefont {Doukas}},
  \bibinfo {author} {\bibfnamefont {W.}~\bibnamefont {Naylor}}, \ and\ \bibinfo
  {author} {\bibfnamefont {A.~S.}\ \bibnamefont {Cornell}},\ }\href {\doibase
  10.1103/PhysRevD.83.124034} {\bibfield  {journal} {\bibinfo  {journal} {Phys.
  Rev.}\ }\textbf {\bibinfo {volume} {D83}},\ \bibinfo {pages} {124034}
  (\bibinfo {year} {2011})},\ \Eprint {http://arxiv.org/abs/1104.1281}
  {arXiv:1104.1281 [hep-th]} \BibitemShut {NoStop}%
\bibitem [{\citenamefont {Dias}\ \emph {et~al.}(2014)\citenamefont {Dias},
  \citenamefont {Hartnett},\ and\ \citenamefont {Santos}}]{Dias:2014eua}%
  \BibitemOpen
  \bibfield  {author} {\bibinfo {author} {\bibfnamefont {s.~J.~C.}\
  \bibnamefont {Dias}}, \bibinfo {author} {\bibfnamefont {G.~S.}\ \bibnamefont
  {Hartnett}}, \ and\ \bibinfo {author} {\bibfnamefont {J.~E.}\ \bibnamefont
  {Santos}},\ }\href {\doibase 10.1088/0264-9381/31/24/245011} {\bibfield
  {journal} {\bibinfo  {journal} {Class. Quant. Grav.}\ }\textbf {\bibinfo
  {volume} {31}},\ \bibinfo {pages} {245011} (\bibinfo {year} {2014})},\
  \Eprint {http://arxiv.org/abs/1402.7047} {arXiv:1402.7047 [hep-th]}
  \BibitemShut {NoStop}%
\bibitem [{\citenamefont {Frolov}\ \emph {et~al.}(2018)\citenamefont {Frolov},
  \citenamefont {Krtouš}, \citenamefont {Kubizňák},\ and\ \citenamefont
  {Santos}}]{Frolov:2018ezx}%
  \BibitemOpen
  \bibfield  {author} {\bibinfo {author} {\bibfnamefont {V.~P.}\ \bibnamefont
  {Frolov}}, \bibinfo {author} {\bibfnamefont {P.}~\bibnamefont {Krtouš}},
  \bibinfo {author} {\bibfnamefont {D.}~\bibnamefont {Kubizňák}}, \ and\
  \bibinfo {author} {\bibfnamefont {J.~E.}\ \bibnamefont {Santos}},\ }\href
  {\doibase 10.1103/PhysRevLett.120.231103} {\bibfield  {journal} {\bibinfo
  {journal} {Phys. Rev. Lett.}\ }\textbf {\bibinfo {volume} {120}},\ \bibinfo
  {pages} {231103} (\bibinfo {year} {2018})},\ \Eprint
  {http://arxiv.org/abs/1804.00030} {arXiv:1804.00030 [hep-th]} \BibitemShut
  {NoStop}%
\bibitem [{\citenamefont {Cho}(2003)}]{Cho:2003qe}%
  \BibitemOpen
  \bibfield  {author} {\bibinfo {author} {\bibfnamefont {H.~T.}\ \bibnamefont
  {Cho}},\ }\href {\doibase 10.1103/PhysRevD.68.024003} {\bibfield  {journal}
  {\bibinfo  {journal} {Phys. Rev.}\ }\textbf {\bibinfo {volume} {D68}},\
  \bibinfo {pages} {024003} (\bibinfo {year} {2003})},\ \Eprint
  {http://arxiv.org/abs/gr-qc/0303078} {arXiv:gr-qc/0303078 [gr-qc]}
  \BibitemShut {NoStop}%
\bibitem [{\citenamefont {Jing}(2005)}]{Jing:2005dt}%
  \BibitemOpen
  \bibfield  {author} {\bibinfo {author} {\bibfnamefont {J.-l.}\ \bibnamefont
  {Jing}},\ }\href {\doibase 10.1103/PhysRevD.71.124006} {\bibfield  {journal}
  {\bibinfo  {journal} {Phys. Rev.}\ }\textbf {\bibinfo {volume} {D71}},\
  \bibinfo {pages} {124006} (\bibinfo {year} {2005})},\ \Eprint
  {http://arxiv.org/abs/gr-qc/0502023} {arXiv:gr-qc/0502023 [gr-qc]}
  \BibitemShut {NoStop}%
\bibitem [{\citenamefont {Chakrabarti}(2009)}]{Chakrabarti:2008xz}%
  \BibitemOpen
  \bibfield  {author} {\bibinfo {author} {\bibfnamefont {S.~K.}\ \bibnamefont
  {Chakrabarti}},\ }\href {\doibase 10.1140/epjc/s10052-009-1026-y} {\bibfield
  {journal} {\bibinfo  {journal} {Eur. Phys. J.}\ }\textbf {\bibinfo {volume}
  {C61}},\ \bibinfo {pages} {477} (\bibinfo {year} {2009})},\ \Eprint
  {http://arxiv.org/abs/0809.1004} {arXiv:0809.1004 [gr-qc]} \BibitemShut
  {NoStop}%
\bibitem [{\citenamefont {Saleh}\ \emph {et~al.}(2016)\citenamefont {Saleh},
  \citenamefont {Bouetou},\ and\ \citenamefont {Kofane}}]{Saleh:2016pke}%
  \BibitemOpen
  \bibfield  {author} {\bibinfo {author} {\bibfnamefont {M.}~\bibnamefont
  {Saleh}}, \bibinfo {author} {\bibfnamefont {B.~T.}\ \bibnamefont {Bouetou}},
  \ and\ \bibinfo {author} {\bibfnamefont {T.~C.}\ \bibnamefont {Kofane}},\
  }\href {\doibase 10.1007/s10509-016-2725-0} {\bibfield  {journal} {\bibinfo
  {journal} {Astrophys. Space Sci.}\ }\textbf {\bibinfo {volume} {361}},\
  \bibinfo {pages} {137} (\bibinfo {year} {2016})},\ \Eprint
  {http://arxiv.org/abs/1604.00820} {arXiv:1604.00820 [gr-qc]} \BibitemShut
  {NoStop}%
\bibitem [{\citenamefont {Fernando}(2015)}]{Fernando:2015hma}%
  \BibitemOpen
  \bibfield  {author} {\bibinfo {author} {\bibfnamefont {S.}~\bibnamefont
  {Fernando}},\ }\href {\doibase 10.1142/S0217732315501473} {\bibfield
  {journal} {\bibinfo  {journal} {Mod. Phys. Lett.}\ }\textbf {\bibinfo
  {volume} {A30}},\ \bibinfo {pages} {1550147} (\bibinfo {year} {2015})},\
  \Eprint {http://arxiv.org/abs/1506.03005} {arXiv:1506.03005 [gr-qc]}
  \BibitemShut {NoStop}%
\bibitem [{\citenamefont {Chowdhury}\ and\ \citenamefont
  {Banerjee}(2018)}]{Chowdhury:2018izv}%
  \BibitemOpen
  \bibfield  {author} {\bibinfo {author} {\bibfnamefont {A.}~\bibnamefont
  {Chowdhury}}\ and\ \bibinfo {author} {\bibfnamefont {N.}~\bibnamefont
  {Banerjee}},\ }\href {\doibase 10.1140/epjc/s10052-018-6065-9} {\bibfield
  {journal} {\bibinfo  {journal} {Eur. Phys. J.}\ }\textbf {\bibinfo {volume}
  {C78}},\ \bibinfo {pages} {594} (\bibinfo {year} {2018})},\ \Eprint
  {http://arxiv.org/abs/1807.09559} {arXiv:1807.09559 [gr-qc]} \BibitemShut
  {NoStop}%
\bibitem [{\citenamefont {Bl\'azquez-Salcedo}\ and\ \citenamefont
  {Knoll}(2018{\natexlab{b}})}]{Blazquez-Salcedo:2018rec}%
  \BibitemOpen
  \bibfield  {author} {\bibinfo {author} {\bibfnamefont {J.~L.}\ \bibnamefont
  {Bl\'azquez-Salcedo}}\ and\ \bibinfo {author} {\bibfnamefont
  {C.}~\bibnamefont {Knoll}},\ }\href@noop {} {\  (\bibinfo {year}
  {2018}{\natexlab{b}})},\ \Eprint {http://arxiv.org/abs/1808.00503}
  {arXiv:1808.00503 [gr-qc]} \BibitemShut {NoStop}%
\bibitem [{\citenamefont {Kerr}(1963)}]{PhysRevLett.11.237}%
  \BibitemOpen
  \bibfield  {author} {\bibinfo {author} {\bibfnamefont {R.~P.}\ \bibnamefont
  {Kerr}},\ }\href {\doibase 10.1103/PhysRevLett.11.237} {\bibfield  {journal}
  {\bibinfo  {journal} {Phys. Rev. Lett.}\ }\textbf {\bibinfo {volume} {11}},\
  \bibinfo {pages} {237} (\bibinfo {year} {1963})}\BibitemShut {NoStop}%
\bibitem [{\citenamefont {Boyer}\ and\ \citenamefont
  {Lindquist}(1967)}]{doi:10.1063/1.1705193}%
  \BibitemOpen
  \bibfield  {author} {\bibinfo {author} {\bibfnamefont {R.~H.}\ \bibnamefont
  {Boyer}}\ and\ \bibinfo {author} {\bibfnamefont {R.~W.}\ \bibnamefont
  {Lindquist}},\ }\href {\doibase 10.1063/1.1705193} {\bibfield  {journal}
  {\bibinfo  {journal} {Journal of Mathematical Physics}\ }\textbf {\bibinfo
  {volume} {8}},\ \bibinfo {pages} {265} (\bibinfo {year} {1967})},\ \Eprint
  {http://arxiv.org/abs/https://doi.org/10.1063/1.1705193}
  {https://doi.org/10.1063/1.1705193} \BibitemShut {NoStop}%
\bibitem [{\citenamefont {Chandrasekhar}(2002)}]{Chandrasekhar:579245}%
  \BibitemOpen
  \bibfield  {author} {\bibinfo {author} {\bibfnamefont {S.}~\bibnamefont
  {Chandrasekhar}},\ }\href {https://cds.cern.ch/record/579245} {\emph
  {\bibinfo {title} {{The mathematical theory of black holes}}}},\ Oxford
  classic texts in the physical sciences\ (\bibinfo  {publisher} {Oxford Univ.
  Press},\ \bibinfo {address} {Oxford},\ \bibinfo {year} {2002})\BibitemShut
  {NoStop}%
\bibitem [{\citenamefont {Roger}()}]{doi:10.1111/j.1749-6632.1973.tb41447.x}%
  \BibitemOpen
  \bibfield  {author} {\bibinfo {author} {\bibfnamefont {P.}~\bibnamefont
  {Roger}},\ }\href {\doibase 10.1111/j.1749-6632.1973.tb41447.x} {\bibfield
  {journal} {\bibinfo  {journal} {Annals of the New York Academy of Sciences}\
  }\textbf {\bibinfo {volume} {224}},\ \bibinfo {pages} {125}},\ \Eprint
  {http://arxiv.org/abs/https://nyaspubs.onlinelibrary.wiley.com/doi/pdf/10.1111/j.1749-6632.1973.tb41447.x}
  {https://nyaspubs.onlinelibrary.wiley.com/doi/pdf/10.1111/j.1749-6632.1973.tb41447.x}
  \BibitemShut {NoStop}%
\bibitem [{\citenamefont {Floyd}(1973)}]{Floyd}%
  \BibitemOpen
  \bibfield  {author} {\bibinfo {author} {\bibfnamefont {R.~M.}\ \bibnamefont
  {Floyd}},\ }\emph {\bibinfo {title} {The dynamics of kerr fields}},\
  \href@noop {} {Ph.D. thesis},\ \bibinfo  {school} {London University},
  \bibinfo {address} {London} (\bibinfo {year} {1973})\BibitemShut {NoStop}%
\bibitem [{\citenamefont {Kalnins}\ and\ \citenamefont
  {Miller}(1992)}]{doi:10.1063/1.529963}%
  \BibitemOpen
  \bibfield  {author} {\bibinfo {author} {\bibfnamefont {E.~G.}\ \bibnamefont
  {Kalnins}}\ and\ \bibinfo {author} {\bibfnamefont {W.}~\bibnamefont
  {Miller}},\ }\href {\doibase 10.1063/1.529963} {\bibfield  {journal}
  {\bibinfo  {journal} {Journal of Mathematical Physics}\ }\textbf {\bibinfo
  {volume} {33}},\ \bibinfo {pages} {286} (\bibinfo {year} {1992})},\ \Eprint
  {http://arxiv.org/abs/https://doi.org/10.1063/1.529963}
  {https://doi.org/10.1063/1.529963} \BibitemShut {NoStop}%
\bibitem [{\citenamefont {Suffern}\ \emph {et~al.}(1983)\citenamefont
  {Suffern}, \citenamefont {Fackerell},\ and\ \citenamefont
  {Cosgrove}}]{doi:10.1063/1.525820}%
  \BibitemOpen
  \bibfield  {author} {\bibinfo {author} {\bibfnamefont {K.~G.}\ \bibnamefont
  {Suffern}}, \bibinfo {author} {\bibfnamefont {E.~D.}\ \bibnamefont
  {Fackerell}}, \ and\ \bibinfo {author} {\bibfnamefont {C.~M.}\ \bibnamefont
  {Cosgrove}},\ }\href {\doibase 10.1063/1.525820} {\bibfield  {journal}
  {\bibinfo  {journal} {Journal of Mathematical Physics}\ }\textbf {\bibinfo
  {volume} {24}},\ \bibinfo {pages} {1350} (\bibinfo {year} {1983})},\ \Eprint
  {http://arxiv.org/abs/https://doi.org/10.1063/1.525820}
  {https://doi.org/10.1063/1.525820} \BibitemShut {NoStop}%
\bibitem [{{\relax DLMF}()}]{NIST:DLMF}%
  \BibitemOpen
  {\relax DLMF},\ \href {http://dlmf.nist.gov/} {\enquote {\bibinfo {title}
  {{\it NIST Digital Library of Mathematical Functions}},}\ }\bibinfo
  {howpublished} {http://dlmf.nist.gov/, Release 1.0.18 of 2018-03-27},\
  \bibinfo {note} {f.~W.~J. Olver, A.~B. {Olde Daalhuis}, D.~W. Lozier, B.~I.
  Schneider, R.~F. Boisvert, C.~W. Clark, B.~R. Miller and B.~V. Saunders,
  eds.}\BibitemShut {Stop}%
\bibitem [{\citenamefont {Leaver}(1985)}]{Leaver:1985ax}%
  \BibitemOpen
  \bibfield  {author} {\bibinfo {author} {\bibfnamefont {E.~W.}\ \bibnamefont
  {Leaver}},\ }\href {\doibase 10.1098/rspa.1985.0119} {\bibfield  {journal}
  {\bibinfo  {journal} {Proc. Roy. Soc. Lond.}\ }\textbf {\bibinfo {volume}
  {A402}},\ \bibinfo {pages} {285} (\bibinfo {year} {1985})}\BibitemShut
  {NoStop}%
\bibitem [{\citenamefont {Myers}\ and\ \citenamefont
  {Perry}(1986)}]{Myers:1986un}%
  \BibitemOpen
  \bibfield  {author} {\bibinfo {author} {\bibfnamefont {R.~C.}\ \bibnamefont
  {Myers}}\ and\ \bibinfo {author} {\bibfnamefont {M.~J.}\ \bibnamefont
  {Perry}},\ }\href {\doibase 10.1016/0003-4916(86)90186-7} {\bibfield
  {journal} {\bibinfo  {journal} {Annals Phys.}\ }\textbf {\bibinfo {volume}
  {172}},\ \bibinfo {pages} {304} (\bibinfo {year} {1986})}\BibitemShut
  {NoStop}%
\bibitem [{\citenamefont {Frolov}\ and\ \citenamefont
  {Kubiznak}(2007)}]{Frolov:2007nt}%
  \BibitemOpen
  \bibfield  {author} {\bibinfo {author} {\bibfnamefont {V.~P.}\ \bibnamefont
  {Frolov}}\ and\ \bibinfo {author} {\bibfnamefont {D.}~\bibnamefont
  {Kubiznak}},\ }\href {\doibase 10.1103/PhysRevLett.98.011101} {\bibfield
  {journal} {\bibinfo  {journal} {Phys. Rev. Lett.}\ }\textbf {\bibinfo
  {volume} {98}},\ \bibinfo {pages} {011101} (\bibinfo {year} {2007})},\
  \Eprint {http://arxiv.org/abs/gr-qc/0605058} {arXiv:gr-qc/0605058 [gr-qc]}
  \BibitemShut {NoStop}%
\bibitem [{\citenamefont {Nollert}(1993)}]{Nollert:1993zz}%
  \BibitemOpen
  \bibfield  {author} {\bibinfo {author} {\bibfnamefont {H.-P.}\ \bibnamefont
  {Nollert}},\ }\href {\doibase 10.1103/PhysRevD.47.5253} {\bibfield  {journal}
  {\bibinfo  {journal} {Phys. Rev.}\ }\textbf {\bibinfo {volume} {D47}},\
  \bibinfo {pages} {5253} (\bibinfo {year} {1993})}\BibitemShut {NoStop}%
\bibitem [{\citenamefont {Andersson}\ and\ \citenamefont
  {Onozawa}(1996)}]{Andersson:1996xw}%
  \BibitemOpen
  \bibfield  {author} {\bibinfo {author} {\bibfnamefont {N.}~\bibnamefont
  {Andersson}}\ and\ \bibinfo {author} {\bibfnamefont {H.}~\bibnamefont
  {Onozawa}},\ }\href {\doibase 10.1103/PhysRevD.54.7470} {\bibfield  {journal}
  {\bibinfo  {journal} {Phys. Rev.}\ }\textbf {\bibinfo {volume} {D54}},\
  \bibinfo {pages} {7470} (\bibinfo {year} {1996})},\ \Eprint
  {http://arxiv.org/abs/gr-qc/9607054} {arXiv:gr-qc/9607054 [gr-qc]}
  \BibitemShut {NoStop}%
\bibitem [{\citenamefont {Cook}\ and\ \citenamefont
  {Zalutskiy}(2016)}]{Cook:2016fge}%
  \BibitemOpen
  \bibfield  {author} {\bibinfo {author} {\bibfnamefont {G.~B.}\ \bibnamefont
  {Cook}}\ and\ \bibinfo {author} {\bibfnamefont {M.}~\bibnamefont
  {Zalutskiy}},\ }\href {\doibase 10.1088/0264-9381/33/24/245008} {\bibfield
  {journal} {\bibinfo  {journal} {Class. Quant. Grav.}\ }\textbf {\bibinfo
  {volume} {33}},\ \bibinfo {pages} {245008} (\bibinfo {year} {2016})},\
  \Eprint {http://arxiv.org/abs/1603.09710} {arXiv:1603.09710 [gr-qc]}
  \BibitemShut {NoStop}%
\end{thebibliography}%

\end{document}